\PassOptionsToPackage{pdfpagelabels=false}{hyperref} 
\documentclass[a4paper,fleqn,usenatbib]{mnras}

\usepackage{newtxtext,newtxmath}

\usepackage[T1]{fontenc}
\usepackage{ae,aecompl}

\usepackage{graphicx}	
\usepackage{amsmath}	
\usepackage{nccmath}
\usepackage{amssymb}	
\usepackage{pdflscape}	
\usepackage{multicol}
\usepackage{afterpage}

\title[SN~2016esw: a luminous Type II supernova]{SN~2016esw: a luminous Type II supernova observed within the first day after the explosion}

\author[T. de Jaeger et al.]{Thomas de Jaeger$^{1}$\thanks{E-mail: tdejaeger@berkeley.edu},
Lluis Galbany$^{2}$, 
Claudia P. Guti\'errez$^{3}$, 
Alexei V. Filippenko$^{1,4}$, 
\newauthor
WeiKang Zheng$^{1}$, 
Thomas G. Brink$^{1}$, 
Ryan~J.~Foley$^{5}$, 
Sebastian F. S\'anchez$^{6}$, 
\newauthor
Sanyum Channa$^{1}$, 
Maxime de Kouchkovsky$^{1}$, 
Goni Halevi$^{1,7}$, 
Charles D. Kilpatrick$^{5}$, 
\newauthor
Sahana Kumar$^{1,8}$, 
Jeffrey Molloy$^{1}$, 
Yen-Chen Pan$^{5}$, 
Timothy W. Ross$^{1}$, 
Isaac Shivvers$^{1}$, 
\newauthor
Matthew R. Siebert$^{5}$, 
Benjamin Stahl$^{1}$, 
Samantha Stegman$^{1}$, 
Sameen Yunus$^{1}$\\
\\
$^{1}$Department of Astronomy, University of California, Berkeley, CA 94720-3411, USA.\\
$^{2}$PITT PACC, Department of Physics and Astronomy, University of Pittsburgh, Pittsburgh, PA 15260, USA.\\
$^{3}$University of Southampton School of Physics and Astronomy, Southampton, SO17 1BJ, UK.\\
$^{4}$Miller Senior Fellow, Miller Institute for Basic Research in Science, University of California, Berkeley, CA 94720, USA.\\
$^{5}$Department of Astronomy and Astrophysics, University of California, Santa Cruz, CA 95064, USA.\\
$^{6}$Instituto de Astronom\'ia,Universidad Nacional Auton\'oma de M\'exico, A.P. 70-264, 04510 M\'exico D.F., Mexico.\\
$^{7}$Department of Astrophysical Sciences, Princeton University, Princeton, NJ 08544, USA.\\
$^{8}$Department of Physics and Astronomy, Florida State University, 32306 USA. 
}
\date{Accepted XXX. Received YYY; in original form ZZZ}

\pubyear{2017}

\begin{document}
\label{firstpage}
\pagerange{\pageref{firstpage}--\pageref{lastpage}}
\maketitle

\begin{abstract}
We present photometry, spectroscopy, and host-galaxy integral-field spectroscopy of the Type II supernova (SN) 2016esw in CGCG~229-009 from the first day after the explosion up to 120 days. Its light-curve shape is similar to that of a typical SN~II; however, SN~2016esw is near the high-luminosity end of the SN~II distribution, with a peak of $M^{\rm max}_{V}=-18.36$ mag. The $V$-band light curve exhibits a long recombination phase for a SN~II (similar to the long-lived plateau of SN~2004et). Considering the well-known relation between the luminosity and the plateau decline rate, SN~2016esw should have a $V$-band slope of $\sim 2.10$ mag (100 days)$^{-1}$; however, SN~2016esw has a substantially flatter plateau with a slope of $1.01\pm 0.26$ mag (100 days)$^{-1}$, perhaps indicating that interacting Type II supernovae are not useful for cosmology. At 19.5 days post-explosion, the spectrum presents a boxy H$\alpha$ emission line with flat absorption profiles, suggesting interaction between the ejecta and circumstellar matter. Finally, based on the spectral properties, SN~2016esw shows similarities with the luminous and interacting SN~2007pk at early epochs, particularly in terms of observable line features and their evolution.

\end{abstract}



\section{Introduction}

Type II supernovae (hereafter SNe~II) are characterised by strong hydrogen Balmer lines in their
spectra. Their progenitors are initially (zero-age main sequence; ZAMS) massive stars ($M_\mathrm{{ZAMS}} \geq 8$ ${\rm M}_{\odot}$; see \citealt{smartt09a} for a review, and references therein) that did not shed their outer envelope prior to exploding.

SNe~II exhibit a large range of observed photometric and spectroscopic properties and can be further separated into several subclassifications (see \citealt{filippenko97}, and references therein). Historically, a first subclassification based on the light-curve morphologies was made: SNe~IIP, which exhibit a long-duration plateau ($\leq 100$~days) of roughly constant luminosity, and SNe~IIL, which show a faster declining light curve \citep{barbon79}. However, recent studies on large datasets (\citealt{anderson14a,sanders15,valenti16,galbany16a}, also see \citealt{rubin16b}) do not favour such bimodality in the diversity, instead suggesting that the SN~IIP and SN~IIL families form a continuous class. Therefore, henceforth we simply refer to these two subgroups as SNe~II. 

Based on the spectroscopic properties, two other subgroups were proposed: SNe~IIb, which evolve spectroscopically from SNe~II at early times to H-deficient a few weeks to a month past maximum brightness \citep{woosley87,filippenko88,filippenko93}, and SNe~IIn, which have relatively narrow H~I emission lines owing to interaction between the ejecta and dense circumstellar matter \citep[CSM;][]{che81,fra82,sch90,filippenko91}. Note that in this work, these last two subgroups are not considered in our photometric or spectroscopic analysis, as their characteristics differ from those of the SN~II family (SNe~IIP and SNe~IIL) to which our object belongs. 

Thanks to direct progenitor detections \citep{vandyk03,smartt09a,smartt09b} and hydrodynamical models \citep{grassberg71,falk77,chevalier76}, it is known that most SN~II progenitors are red supergiant stars (RSGs; 8--16 ${\rm M}_{\odot}$). Nevertheless, we see a great variety of SNe~II in the Universe \citep[e.g.,][]{filippenko91,hamuy03a,anderson14a,gutierrez17a}. This diversity results from differences among progenitors and explosion mechanisms, such as the mass of the H envelope, the radius, the metallicity, the explosion energy, the mass loss, and the amount and nature of interaction with CSM. For example, faster declining SN~II progenitor (historically SNe~IIL) may have greater ZAMS masses \citep{EliasRosa10,EliasRosa11} and smaller H envelopes \citep{popov93}, or be RSGs surrounded by dense CSM \citep{morozova17}.

As SNe~II have also been established to be useful independent distance indicators \citep{hamuy02}, a better understanding of SN~II diversity will be useful not only for our knowledge of SN~II properties but also for reducing the scatter in the Hubble diagram, and thus obtaining tighter constraints on the cosmological parameters \citep{dejaeger17a,dejaeger17b}.

SN~2016esw provides a good opportunity to explore the zoo of SNe~II owing to the acquisition of well-sampled data starting less than one day after the explosion. In this paper, we present optical photometry and spectroscopy of this object, as well as Potsdam Multi Aperture Spectrograph \citep[PMAS;][]{roth05} integral-field unit (IFU) spectroscopy of its host galaxy. We show that SN~2016esw is part of the brightest tail of the SN~II luminosity distribution (excluding SNe~IIn or superluminous SNe~II) and also exhibits some signs of interaction with CSM. Given the observed characteristics of the transient, we discuss them in comparison to other SNe~II with the aim of defining its nature and constraining its physical properties.

The paper is organised as follows. Section 2 contains a description of the observations and the data. In Section 3, we describe the evolution of the light curves, colour curves, and spectra, and we estimate the reddening. We discuss our results in Section 4, and we conclude with a summary in Section 5.

\section{Observations}

\subsection{Discovery}

\begin{figure}
\includegraphics[width=8cm,height=5cm]{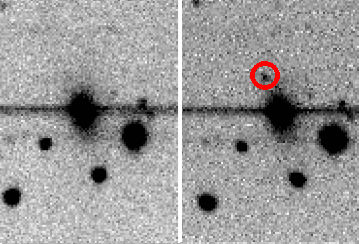}
 \caption{\textit{Left:} Pre-explosion unfiltered image taken on 2016 August 7.31 with KAIT at Lick Observatory. \textit{Right:} KAIT discovery unfiltered image taken on 2016 August 8.31. SN~2016esw is marked with the red circle. The horizontal streaks are artifacts produced by saturated stars.}
 \label{fig:pre_explosion}
\end{figure}

As part of the Lick Observatory Supernova Search \citep[LOSS]{filippenko01,leaman11}, SN~2016esw was discovered by \citet{haveli16} in an unfiltered\footnote{The unfiltered band is similar to the $R$ band; see \citealt{li03}} image (unfiltered apparent magnitude $\sim 18.5$) on 2016 August 8.31 (UT dates are used throughout this paper; JD = 2,457,608.81) with the 0.76~m Katzman Automatic Imaging Telescope (KAIT). The transient was located at $\alpha = 18^{\rm h}58^{\rm m}04\fs09 \pm 0\fs1$, $\delta = +43^{\circ}56'22{\farcs}60 \pm 0{\farcs}5$ (J2000.0), and it is associated with the galaxy CGCG~229-009\footnote{Also called 2MASX J18580350+4356090.} at redshift $z=0.02831$ (see Section \ref{sec:host}). With respect to the galaxy nucleus, the object is situated 6{\farcs}85 east and 13{\farcs}45 north. \citet{shivvers16} classified the object as a SNe~II owing to the presence of a blue continuum with weak, broad H Balmer lines. The most recent pre-explosion nondetection in archival images was obtained with LOSS/KAIT on 2016 August 7.31 (limiting magnitude 19.3; JD $=$ 2,457,607.81), providing very good constraints on the explosion time (see Figure \ref{fig:pre_explosion}). 

To estimate a more accurate and precise explosion date than the midpoint (JD = 2,457,608.31 $\pm$ 0.50) between the discovery and the last nondetection epochs, the early-time unfiltered light curve (the data prior to maximum brightness and including nondetections) is fitted using a simple power law \citep{gonzalezgaitan14}. We obtain an explosion date ($T_{\rm exp}$) of JD = 2,457,608.33 $\pm$ 0.15 (2016 August 7.83). All observations are presented with respect to this explosion date.

\subsection{Photometry}

\begin{figure}
	\includegraphics[width=8cm,height=8cm]{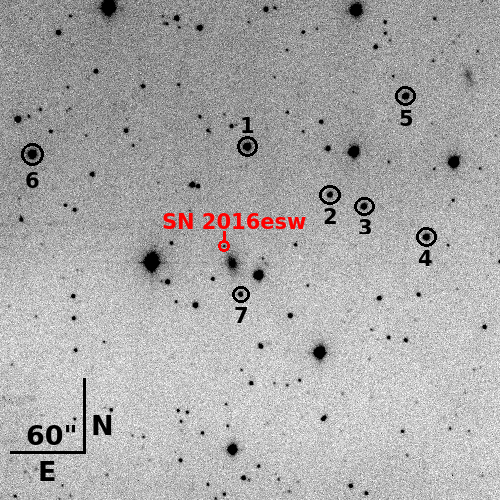}
 \caption{KAIT $I$-band image showing the location of SN~2016esw with a red circle. Reference stars are also marked with black circles.}
 \label{fig:Finding_chart}
\end{figure}

Multiband observations ($BVRI$) started only $\sim 2$~hr after the discovery, on 2016 August 8.39 with KAIT and the 1~m Nickel reflector at Lick Observatory. Automated reductions (bias and flat-field corrections and astrometric solution) had been applied to all the images using our image-reduction pipeline \citep{ganeshalingam10}. Even though the transient is relatively distant from its host (see Figure \ref{fig:Finding_chart}), we took care to correctly remove the underlying host-galaxy light. The templates used for final subtractions were taken $\sim 450$ days after the explosion and were geometrically transformed to each individual science frame. 

\begin{figure}
	\includegraphics[width=9cm,height=9cm]{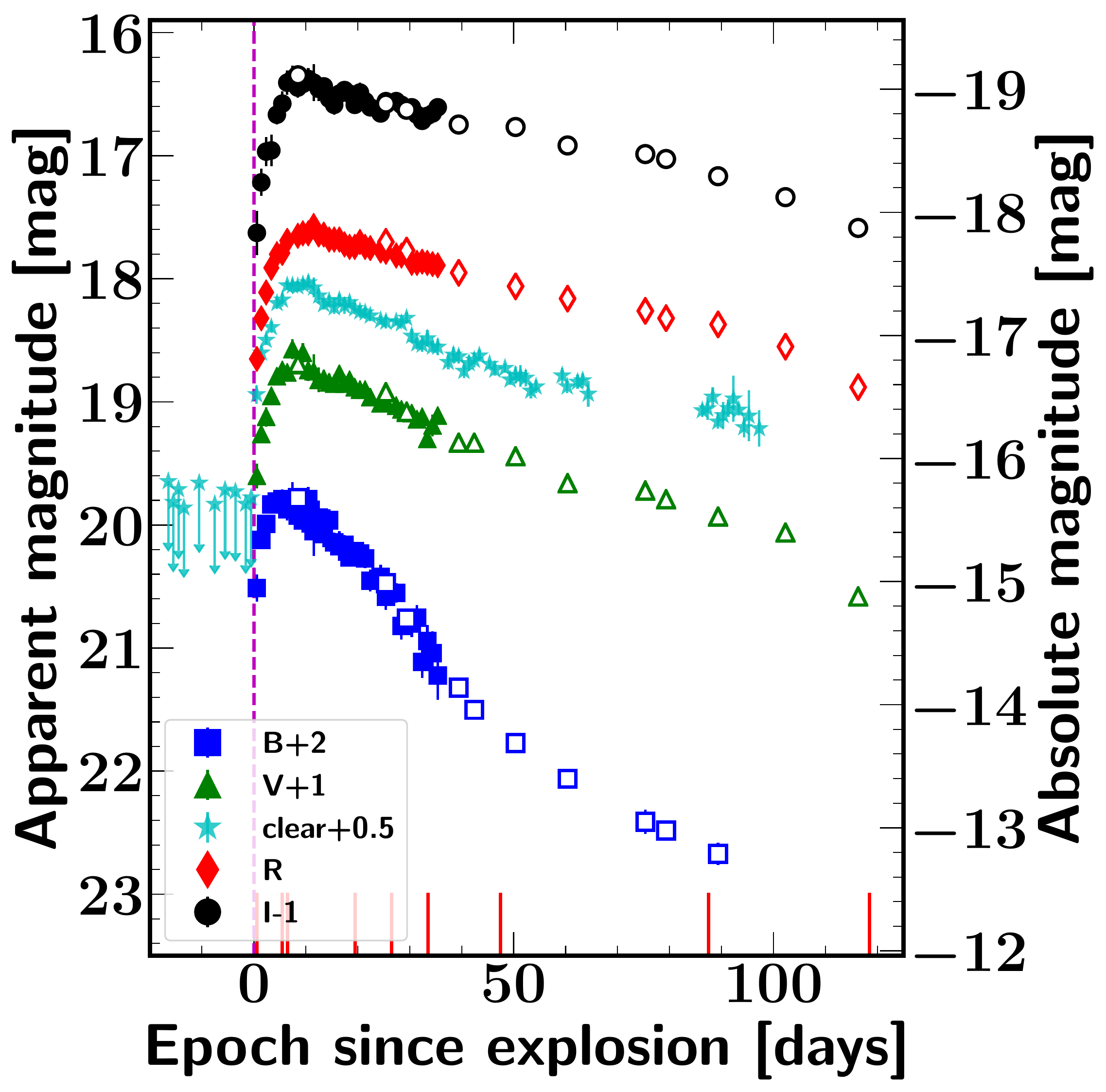}
\caption{SN~2016esw $B$, $V$, $R$, $I$, and unfiltered light curves are shown with blue squares, green triangles, red diamonds, black dots, and cyan stars (respectively). Full and empty symbols represent the data obtained using KAIT and the Nickel telescope, respectively. Cyan arrows show the nondetections. Epochs are with respect to the explosion date (2016 August 7.81 = 0 day), which is represented by the dashed magenta vertical line. Vertical red lines represent epochs of optical spectroscopy. The apparent magnitude is shown on the left ordinate axis, while the absolute magnitude (calculated using only the distance modulus) is represented on the right. The photometry is corrected for Milky Way extinction assuming a \citet{car89} law with $R_V=3.1$, and a dust extinction in the $V$ band of 0.188 mag \citep{schlafly11}.}

 \label{fig:LC_sn2016esw}
\end{figure}

Point-spread-function (PSF) photometry was performed using DAOPHOT \citep{stetson87} from the IDL Astronomy User's Library. Instrumental magnitudes were calibrated relative to a sequence of 7 stars in the field of CGCG~229-009 from the AAVSO Photometric All-Sky Survey (APASS). The magnitudes were then transformed into the Landolt standard system \citep{landolt92} using the empirical prescription presented by Robert Lupton\footnote{\url{http://classic.sdss.org/dr7/algorithms/sdssUBVRITransform.html}}. The resulting $B$, $V$, $R$, and $I$ magnitudes of the local sequence stars are reported in Table \ref{tab:local_seq}, while the photometry for SN~2016esw is presented in Table \ref{tab:sn16esw_photo}. Finally, $BVRI$ light curves corrected for Milky Way Galaxy (MWG) extinction are displayed in Figure \ref{fig:LC_sn2016esw}.

\begin{table*}
\begin{minipage}{166mm}
	\centering

	\caption{$B$, $V$, $R$, and $I$ Local Sequence Star Magnitudes.}
	\begin{tabular}{lccccccr} 
		\hline
		Star & $\alpha$ ($1\sigma$) & $\delta$ ($1\sigma$) & $B$ ($1\sigma$) & $V$ ($1\sigma$) & $R$ ($1\sigma$) & $I$ ($1\sigma$)\\
		& $^{\circ}$ (\arcsec) & $^{\circ}$ (\arcsec) & mag (mag) & mag (mag) & mag (mag) & mag (mag)\\
		\hline
		\hline
1	&284.510  (0.350)  &43.962  (0.302)   &14.365  (0.020)  &13.715  (0.028)  &13.410  (0.102)  &13.032  (0.106)\\
2	&284.484  (0.488)  &43.951  (0.700)   &16.045  (0.025)  &15.270  (0.067)  &14.860  (0.116)  &14.466  (0.189)\\
3	&284.474  (0.506)  &43.948  (0.390)   &15.957  (0.042)  &15.000  (0.040)  &14.507  (0.126)  &14.013  (0.184)\\
4	&284.455  (0.347)  &43.942  (0.519)   &15.958  (0.044)  &14.829  (0.019)  &14.210  (0.125)  &13.648  (0.127)\\
5	&284.462  (0.438)  &43.973  (0.221)   &15.201  (0.022)  &14.487  (0.031)  &14.122  (0.146)  &13.728  (0.090)\\
6	&284.576  (0.215)  &43.959  (0.427)   &15.099  (0.026)  &13.869  (0.048)  &13.241  (0.068)  &12.639  (0.072)\\
7	&284.497  (0.585)  &43.924  (0.445)   &16.878  (0.008)  &15.952  (0.014)  &15.490  (0.139)  &14.998  (0.051)\\
		\hline

\end{tabular}
\label{tab:local_seq}
\end{minipage}
\end{table*}

\begin{table*}
\begin{minipage}{156mm}
	\centering
	\caption{$B$, $V$, $R$, and $I$ Optical Photometry of SN 2016esw.}
	\begin{tabular}{lcccccr} 
		\hline 
		JD$-$2,400,000.5 & Epoch & $B$ ($1\sigma$) & $V$ ($1\sigma$) & $R$ ($1\sigma$) & $I$ ($1\sigma$) &Telescope\\
		&  days  & mag (mag) & mag (mag) & mag (mag) & mag (mag) &\\
		\hline
		\hline
57608.3945 & 0.6 & 18.76 (0.11) & 18.79 (0.10) & 18.80 (0.08) & 18.73 (0.18) & KAIT \\ 
57609.2461 & 1.4 & 18.37 (0.06) & 18.45 (0.06) & 18.47 (0.06) & 18.32 (0.11) & KAIT \\ 
57610.1797 & 2.4 & 18.24 (0.07) & 18.31 (0.08) & 18.26 (0.06) & 18.07 (0.12) & KAIT \\ 
57611.1797 & 3.4 & 18.08 (0.07) & 18.14 (0.06) & 18.06 (0.04) & 18.06 (0.13) & KAIT \\ 
57612.2540 & 4.4 & 18.06 (0.06) & 17.98 (0.06) & 17.95 (0.06) & 17.77 (0.08) & KAIT \\ 
57613.2930 & 5.5 & 18.04 (0.08) & 17.93 (0.05) & 17.94 (0.06) & 17.68 (0.10) & KAIT \\ 
57614.2422 & 6.4 & 18.12 (0.09) & 17.95 (0.06) & 17.84 (0.05) & 17.51 (0.10) & KAIT \\ 
57615.2422 & 7.4 & 18.04 (0.14) & 17.76 (0.08) & $\cdots$ ($\cdots$) & 17.48 (0.11) & KAIT \\ 
57616.2734 & 8.5 & 18.17 (0.10) & 17.89 (0.05) & 17.80 (0.05) & 17.55 (0.08) & KAIT \\ 
57616.3047 & 8.5 & 18.03 (0.01) & 17.88 (0.01) & $\cdots$ ($\cdots$) & 17.45 (0.04) & Nickel \\ 
57617.2461 & 9.4 & 18.21 (0.10) & 17.79 (0.08) & 17.78 (0.06) & 17.52 (0.08) & KAIT \\ 
57618.2773 & 10.5 & 18.04 (0.10) & 17.93 (0.09) & 17.78 (0.07) & 17.48 (0.09) & KAIT \\ 
57619.3242 & 11.5 & 18.30 (0.20) & 17.94 (0.14) & 17.72 (0.09) & 17.51 (0.15) & KAIT \\ 
57620.2891 & 12.5 & 18.19 (0.14) & 18.01 (0.10) & 17.79 (0.07) & 17.56 (0.10) & KAIT \\ 
57621.2734 & 13.5 & 18.32 (0.11) & 18.00 (0.07) & 17.79 (0.05) & 17.54 (0.07) & KAIT \\ 
57622.3008 & 14.5 & 18.21 (0.09) & 18.03 (0.07) & 17.83 (0.07) & 17.64 (0.08) & KAIT \\ 
57623.2734 & 15.5 & 18.39 (0.08) & 18.04 (0.06) & 17.83 (0.05) & 17.69 (0.08) & KAIT \\ 
57624.2656 & 16.5 & 18.42 (0.12) & 17.96 (0.06) & 17.83 (0.06) & 17.60 (0.07) & KAIT \\ 
57625.2578 & 17.4 & 18.41 (0.08) & 18.04 (0.06) & 17.87 (0.04) & 17.57 (0.07) & KAIT \\ 
57626.2344 & 18.4 & 18.51 (0.07) & 18.01 (0.05) & 17.89 (0.05) & 17.60 (0.08) & KAIT \\ 
57627.2305 & 19.4 & 18.46 (0.07) & 18.07 (0.04) & 17.89 (0.05) & 17.69 (0.07) & KAIT \\ 
57628.2305 & 20.4 & 18.48 (0.07) & 18.09 (0.07) & 17.85 (0.04) & 17.59 (0.08) & KAIT \\ 
57629.2227 & 21.4 & 18.52 (0.08) & 18.09 (0.05) & 17.89 (0.04) & 17.66 (0.06) & KAIT \\ 
57630.2227 & 22.4 & 18.70 (0.09) & 18.15 (0.06) & 17.90 (0.05) & 17.71 (0.07) & KAIT \\ 
57632.2109 & 24.4 & 18.67 (0.10) & 18.20 (0.06) & 17.92 (0.04) & 17.76 (0.07) & KAIT \\ 
57633.2188 & 25.4 & 18.72 (0.02) & 18.11 (0.01) & 17.85 (0.01) & 17.68 (0.03) & Nickel \\ 
57633.2266 & 25.4 & 18.83 (0.11) & 18.19 (0.05) & 17.93 (0.05) & 17.66 (0.06) & KAIT \\ 
57635.2070 & 27.4 & 18.80 (0.08) & 18.22 (0.05) & 17.96 (0.04) & 17.66 (0.07) & KAIT \\ 
57636.2148 & 28.4 & 19.07 (0.11) & 18.25 (0.05) & 17.96 (0.04) & 17.69 (0.06) & KAIT \\ 
57637.2109 & 29.4 & 19.01 (0.02) & 18.27 (0.02) & 17.92 (0.01) & 17.73 (0.02) & Nickel \\ 
57638.1953 & 30.4 & 19.05 (0.11) & 18.28 (0.05) & 18.02 (0.05) & 17.71 (0.06) & KAIT \\ 
57639.2031 & 31.4 & 19.00 (0.10) & 18.33 (0.05) & 18.02 (0.05) & 17.77 (0.07) & KAIT \\ 
57640.2031 & 32.4 & 19.36 (0.13) & 18.31 (0.06) & 18.01 (0.05) & 17.82 (0.07) & KAIT \\ 
57641.1875 & 33.4 & 19.19 (0.13) & 18.49 (0.07) & 18.02 (0.05) & 17.78 (0.07) & KAIT \\ 
57642.2031 & 34.4 & 19.29 (0.18) & 18.38 (0.09) & 18.03 (0.06) & 17.76 (0.08) & KAIT \\ 
57643.2031 & 35.4 & 19.47 (0.20) & 18.30 (0.06) & 18.04 (0.05) & 17.71 (0.07) & KAIT \\ 
57647.2305 & 39.4 & 19.57 (0.05) & 18.52 (0.03) & 18.10 (0.02) & 17.85 (0.03) & Nickel \\ 
57650.2227 & 42.4 & 19.75 (0.05) & 18.52 (0.03) & $\cdots$ ($\cdots$) & $\cdots$ ($\cdots$) & Nickel \\ 
57658.1797 & 50.4 & 20.02 (0.05) & 18.63 (0.02) & 18.21 (0.02) & 17.87 (0.03) & Nickel \\ 
57668.1836 & 60.4 & 20.31 (0.07) & 18.85 (0.03) & 18.31 (0.02) & 18.02 (0.03) & Nickel \\ 
57683.1445 & 75.3 & 20.66 (0.10) & 18.91 (0.03) & 18.41 (0.02) & 18.09 (0.03) & Nickel \\ 
57687.1367 & 79.3 & 20.73 (0.08) & 18.98 (0.03) & 18.47 (0.04) & 18.13 (0.03) & Nickel \\ 
57697.1445 & 89.3 & 20.92 (0.09) & 19.12 (0.03) & 18.52 (0.02) & 18.27 (0.04) & Nickel \\ 
57710.1133 & 102.3 & $\cdots$ ($\cdots$) & 19.25 (0.04) & 18.70 (0.03) & 18.44 (0.03) & Nickel \\ 
57724.0977 & 116.3 & $\cdots$ ($\cdots$) & 19.77 (0.06) & 19.03 (0.05) & 18.69 (0.05) & Nickel \\ 
		\hline
\end{tabular}

\label{tab:sn16esw_photo}
\end{minipage}
\end{table*}

\subsection{Optical Spectroscopy}

Spectroscopy of SN 2016esw began on 2016 August 8.41 ($\sim 0.6$ day after explosion) using the Kitt Peak Ohio State Multi-Object Spectrograph (KOSMOS; \citealt{martini14}) mounted on the Mayall 4~m telescope at Kitt Peak National Observatory (KPNO). After this first observation, an intensive spectroscopic follow-up campaign started using the Kast double spectrograph \citep{miller93} on the Shane 3~m telescope at Lick Observatory. To minimise slit losses from atmospheric dispersion, all of the spectra were taken at the parallactic angle \citep{filippenko82}. In total, we obtained nine optical spectra covering the wavelength range from $\sim 3400$ to 10,500 \AA\ with a resolution of $\sim 10$~\AA. Owing to visibility constraints, follow-up spectroscopy stopped on 2016 December 04.13, $\sim 118.5$ days after the explosion. A detailed log of our optical observations is given in Table \ref{tab:log_spectra}.

Spectroscopic reductions were performed using standard IRAF\footnote{IRAF is distributed by the National Optical Astronomy Observatory, which is operated by AURA, Inc., under a cooperative agreement with the U.S. National Science Foundation (NSF).} routines. All data were debiased, flat-fielded, and cleaned of cosmic rays. Extracted one-dimensional spectra were wavelength calibrated using Ar, He, Hg-Cs, Hg-Ar, and Ne lamps. Flux calibration was determined with instrumental sensitivity curves of spectrophotometric standard stars observed on the same night and with the same instrumental setup. Absolute flux calibration was revised using multicolour photometry, integrating the spectral flux transmitted by standard $BVRI$ filters and adjusting it by a multiplicative factor. To ensure that the resulting flux calibration is accurate to within $\sim 0.1$ mag, flux calibration is performed using an iterative procedure. Atmospheric (telluric) absorption features were removed using the well-exposed spectrum of a standard star.

\begin{figure*}
	\includegraphics[width=17cm]{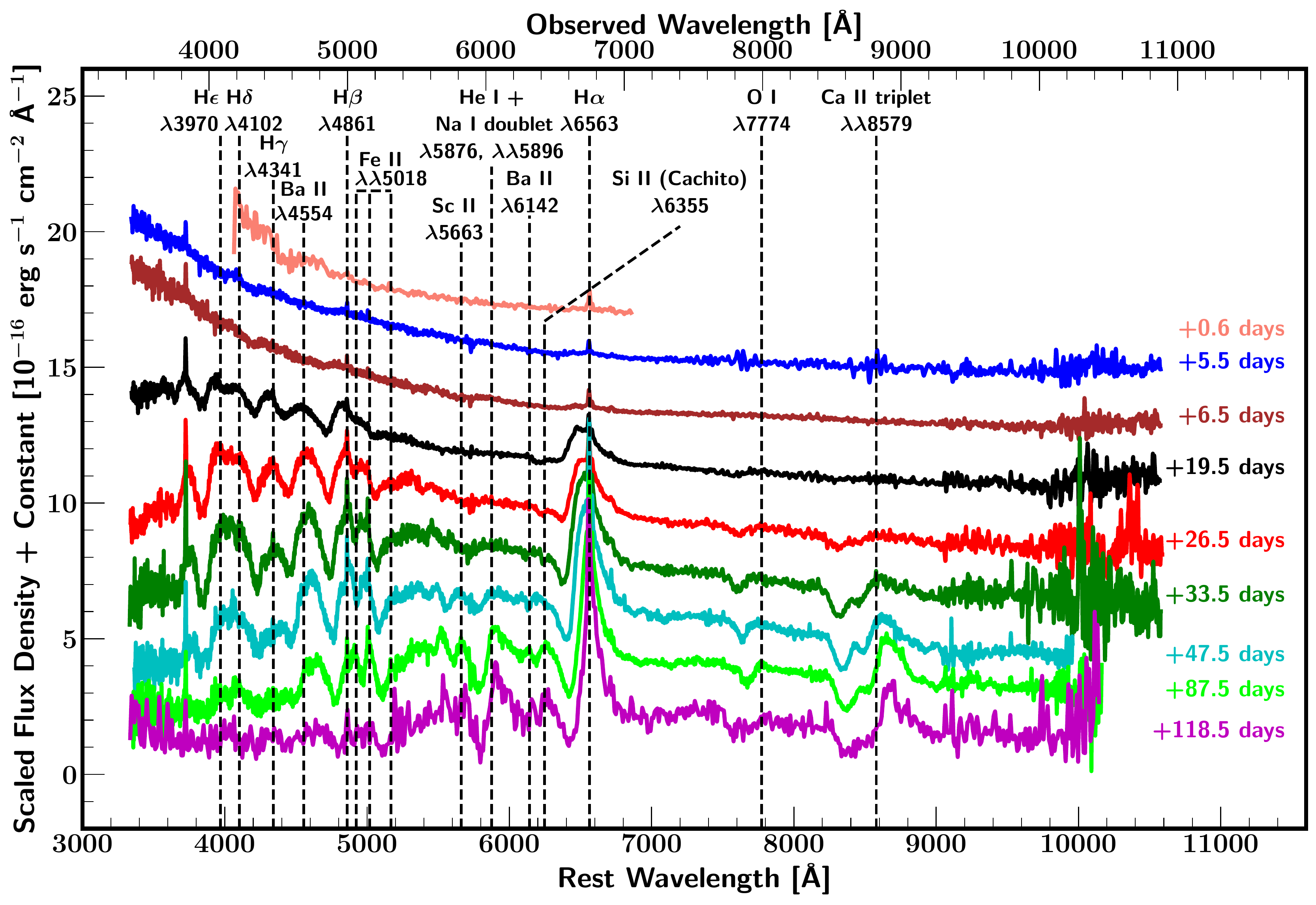}
	\center
 \caption{Optical spectra of SN 2016esw taken with the KPNO/Mayall and Lick/Shane telescopes. On the right, the epochs are listed with respect to the explosion (2016 August 7.81 = 0 day). Telluric absorption features were removed. The flux calibration of the spectra was revised using multicolour photometry. The spectra have been corrected for MWG extinction but not for host-galaxy reddening.}
\label{fig:spectral_sequence}
\end{figure*}

The spectra were corrected for MWG extinction assuming a \citet{car89} law with $R_V=3.1$, and a dust extinction in the $V$ band of 0.188 mag \citep{schlafly11}. However, we do not correct for host-galaxy extinction ($A_{\rm Vh}$) here, as its value is discussed in Section \ref{txt:avh}. The final spectral sequence is displayed in Figure \ref{fig:spectral_sequence}. All of the spectra are available for download from the Berkeley SuperNova DataBase (SNDB\footnote{\url{http://heracles.astro.berkeley.edu/sndb/}}; \citealt{silverman12}).

From the IFU spectroscopy of the SN~2016esw host galaxy and using the forbidden ionised nitrogen doublet [N~II] ($\lambda\lambda$6583, 6548) together with the strong H$\alpha$ emission lines ($\lambda$6563), we measure a redshift of $z = 0.02825 \pm 0.0002$, consistent with the published redshift (0.02831; \citealt{shivvers16}). In the rest of the paper, we adopt the redshift derived from the IFU spectroscopy observation.

\begin{table*}
\begin{minipage}{0.80\linewidth}
	\caption{Log of Optical Spectroscopy of SN~2016esw.}
	\begin{tabular}{lccccccr} 
		\hline
		UT Date & Epoch & Tel./Instru & Wavelength & Resolution &Exp. &Observer &Reducer\\
		 & &(days) & (\AA) & (\AA) & (s) & &\\
		\hline
		2016-08-08.41 &0.6 & Mayall/Kosmos & 4180--7060 & 3 &1200 &CK \& MS &YCP\\
		2016-08-13.28 &5.5 & Shane/Kast & 3438--10,876 & 10 &3600 &IS &HY\\
		2016-08-14.29 &6.5 & Shane/Kast & 3440--10,876 & 10 &3600 &WZ &HY\\
		2016-08-27.22 &19.5 & Shane/Kast & 3440--10,850 & 10 &5400 &PK &HY\\
		2016-09-03.22 &26.5 & Shane/Kast & 3432--10,854 & 10 &3600 &BS &HY\\
		2016-09-10.20 &33.5 & Shane/Kast & 3426--10,878 & 10 &3600 &GH &HY\\
		2016-09-24.16 &47.5 & Shane/Kast & 3460--10,240 & 10 &3600 &WZ &IS\\
		2016-11-03.16 &87.5 & Shane/Kast & 3440--10,450 & 10 &3660 &TB,IS &TB\\
		2016-12-04.13 &118.5 & Shane/Kast & 3430--10,274 & 10 &3000 &TB \& TdJ &TB\\
		\hline
\end{tabular}
Note: Column 1: UT observation date. Column 2: epoch after explosion in days. Column 3: the telescope and instrument used to obtain the spectrum. Columns 4, 5, and 6: wavelength range (\AA), resolution (\AA), and exposure time (s), respectively. Columns 7 and 8: the observers and data reducers are indicated with their initials: CK, Charles Kilpatrick; MS, Matthew Siebert; YCP, Yen-Chen Pan; IS, Isaac Shivvers; HY, Heechan Yuk; WZ, WeiKang Zheng; PK, Patrick Kelly; GH, Goni Halevi; BS, Benjamin Stahl; TB, Thomas Brink; TdJ, Thomas de Jaeger.
\label{tab:log_spectra}
\end{minipage}
\end{table*}

\subsection{Host Galaxy}\label{sec:host}

We obtained integral field spectroscopy (IFS) of the SN~2016esw host galaxy CGCG 229-009 ($3 \times 900$~s, where the second and third exposures are taken with an offset) with the Potsdam Multi Aperture Spectrograph \citep[PMAS;][]{roth05} in PPAK (331 science fibres and 36 for sky measurements) mode \citep{verheijen04} mounted on the 3.5~m telescope of the Centro Astron\'omico Hispano Alem\'an (CAHA) located at Calar Alto Observatory in Almer\'ia, Spain. The data were taken on 2017 August 18 (JD = 2,457,938), about 375 days after the explosion. At that time, the SN was almost undetectable and therefore no strong contamination is expected in the IFS analysed data. The IFS is an asset, providing additional information about both the SN position and those of the entire host galaxy. The observations are included in the PMAS/PPAK Integral-field Supernova hosts COmpilation (PISCO; \citealt{galbany18}), a sample of SN host galaxies observed with the same instrument.

\begin{figure}
\begin{center}
\includegraphics[width=\columnwidth]{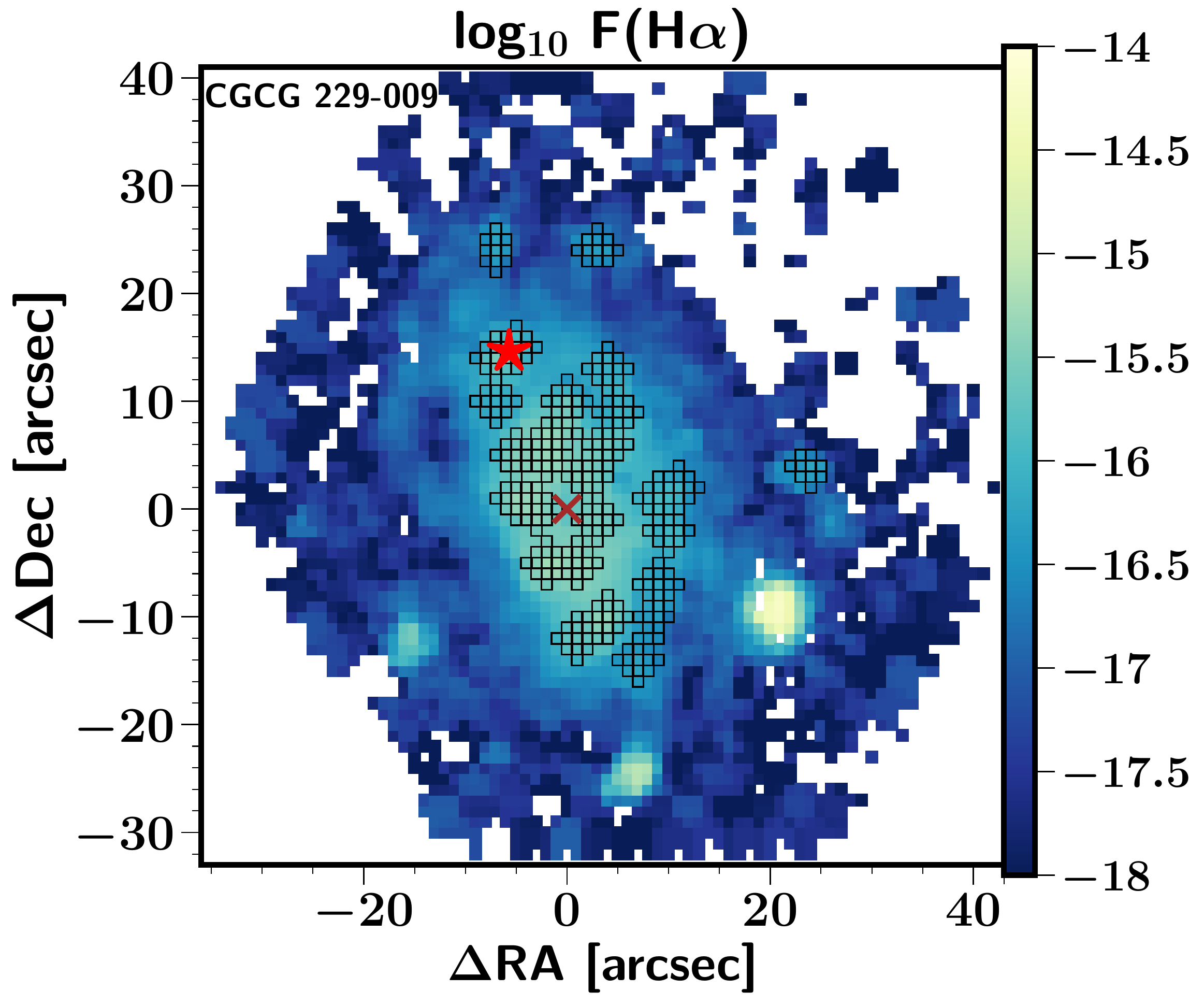}
\caption{Continuum-subtracted H$\alpha$ emission map of CGCG 229-009. The transparent black squares respresent the H~II regions extracted with HIIexplorer. The red star and brown cross represent (respectively) the location of SN~2016esw and the CGCG 229-009 nucleus. The three bumps (white) in the southern part of the galaxy are field stars and, therefore, not selected in our H~II region analysis.}
\label{fig:host}
\end{center}
\end{figure}

The description of the data reduction is presented by \citet{galbany17}, and our analysis follows that presented by \citet{galbany14,galbany16b}. In brief, we perform stellar population synthesis by fitting a base of simple stellar population (SSP) models to all $\sim 4000$ individual spectra using STARLIGHT \citep{cidfernandes05}. The best combination of SSPs is subtracted from the observed spectra to obtain a pure emission-line spectrum, and we accurately measure the flux of the most prominent continuum-free emission lines by means of weighted nonlinear least-squares fits with a single Gaussian plus a linear term.
From the H$\alpha$ emission two-dimensional (2D) map, we select 25 H~II regions by segregating clumps of emission using HIIexplorer\footnote{\url{http://www.caha.es/sanchez/HII_explorer/}} (see \citealt{sanchez12} and \citealt{galbany18} for criteria selection). The SN ``parent'' H~II region is selected as the closest to the SN~2016esw position (see Figure \ref{fig:host}). Note that, owing to the angular resolution of 1\arcsec~(i.e., a spatial resolution of $\sim 560$ pc), our selected H~II regions can actually be aggregations of 1--6 classical H~II regions \citep{mast14}.
Finally, to estimate the galaxy global properties, we sum up all spectra in the 3D datacube with signal-to-noise ratio larger than 1. The resulting total spectrum is analysed following the same steps.

\section{Results}

Here we first describe the photometric data and analyse the light curves, the colours, and the host-galaxy extinction. Then, we discuss the spectroscopic data --- including the spectral evolution, the expansion velocities, and the temperature. Throughout this section, the photometry and the spectra are solely corrected for MWG extinction assuming a \citet{car89} law with $R_V=3.1$, and a dust extinction in the $V$ band of 0.188 mag \citep{schlafly11}. Then, in Section \ref{txt:avh}, we discuss the value of the host-galaxy reddening ($A_{\rm Vh}$) and use it in Section \ref{txt:discussion} to compare with other SNe~II from the literature.

\subsection{Photometry}

\subsubsection{Optical Light Curves}

$B$, $V$, $R$, $I$, and unfiltered light curves of SN~2016esw are plotted in Figure \ref{fig:LC_sn2016esw}. The SN~2016esw light curves show, as is typical for SN~II light curves \citep{anderson14a,sanders15,valenti16,galbany16a}, a fast rise to peak magnitude, followed by a recombination phase where the hydrogen previously ionised by the SN shock recombines. The final SN~II light-curve phase, called the radioactive tail (powered by the conversion of $^{56}$Co into $^{56}$Fe), is not seen as the object was not observable from Lick Observatory at later epochs. The start of the drop into the radioactive tail is clearly visible in the $V$ band between the penultimate and the last photometric points. If the last point belonged to the plateau phase, it should have a magnitude of 19.39 (the plateau decline rate is 1.01 mag (100 days)$^{-1}$; see below). However, the last point has a magnitude of $19.77 \pm 0.06$, or $\sim 6\sigma$ fainter than the plateau brightness at this epoch. 

Note the possible presence of a double-peak structure around day 10 in the $V$ band (perhaps also in the $I$ and unfiltered bands), which has been previously seen in some other SNe~II (e.g., SN~1999em; \citealt{leonard02}). If this structure were real, it could be attributed to the appearance of \ion{Fe}{ii} absorption lines, which mark the onset of the recombination phase \citep{leonard02}. However, for SN~2016esw, we believe that the double peak is not real, as it is not seen in all of the bands; for example, we see it in the unfiltered band but not in the $R$ band, yet both bands are similar \citep{li03}. Finally, we also believe that the feature seen in the $I$ band $\sim 35$ days after the explosion is not real, as it is not seen in the other bands.

\begin{figure}
\begin{center}
\includegraphics[width=\columnwidth]{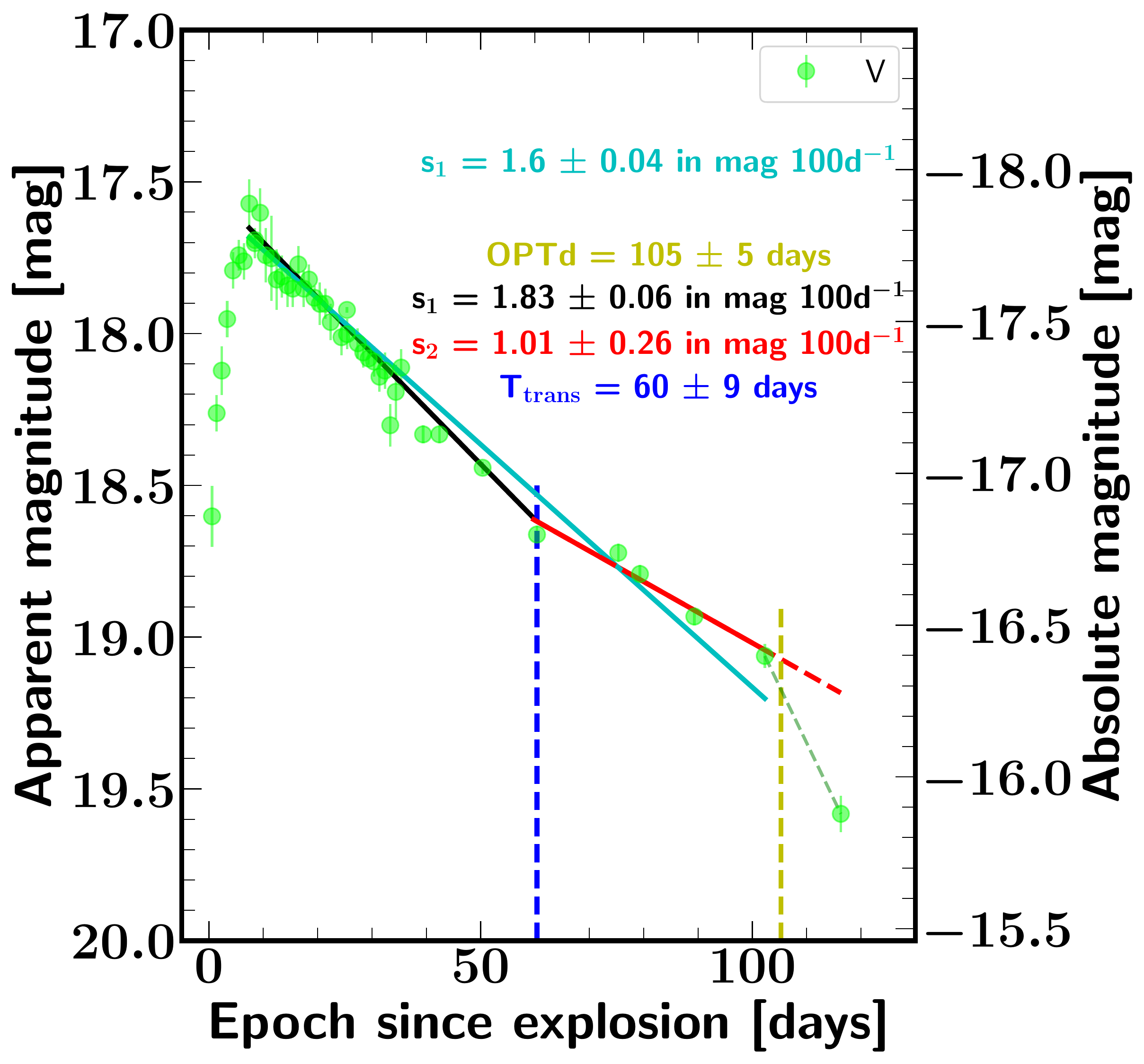}
\caption{$V$-band light-curve fitting following the \citet{anderson14a} procedure. Cyan line is obtained using only one slope while the lines in black and red is using two slopes. The photometry is only corrected for Milky Way extinction.}
\label{fig:lc_fit}
\end{center}
\end{figure}

To define the nature of the transient and to constrain its properties, here we discuss in more detail the characteristics of the $V$-band light curve, which is the band used by \citet{anderson14a}. For this purpose, following their procedure, we perform a $V$-band light-curve fitting using a linear model with four parameters: the two decline rates ($s_{1}$ and $s_{2}$), the epoch of transition between the two slopes ($t_{\rm trans}$), and the magnitude offset. Then, to choose between one slope (only $s_{1}$) or two slopes ($s_{1}$ and $s_{2}$), an $F$-test is performed. Note that we only fit the optically thick phase --- the epoch from maximum brightness to the end of the plateau. Figure \ref{fig:lc_fit} shows the $V$-band light-curve fitting with the two slopes.

After the explosion, the $V$-band brightness rises to its maximum in $\sim 7$ days ($\sim 8$ days in the $R$ band), consistent with the average values found in the literature \citep{inserra12,gonzalezgaitan14,gall15}. The maximum apparent magnitude ($m^{\rm max}_V$) is $17.67 \pm 0.06$ mag, corresponding to an absolute magnitude $M^{\rm max}_V = -17.79 \pm 0.06$ mag (see Section \ref{txt:avh} for more details). The magnitude at maximum is measured by fitting a low-order polynomial to the photometry around the brightest photometric point ($\pm 5$ days).

Following maximum brightness, an initial decline ($s_{1}$) lasting $60 \pm 9$ days at a rate of $1.83 \pm 0.06$ mag (100 days)$^{-1}$ is seen. After the initial cooling, the object enters the plateau phase ($s_{2}$) and the brightness declines more slowly, with a slope of $1.01\pm 0.26$ mag (100 days)$^{-1}$ until $105 \pm 5$ days after the explosion. All these values are consistent with those derived by \citet{anderson14a} ($s_{1} = 2.65 \pm 1.50$, $s_{2} = 1.27 \pm 0.93$, and OPTd = $84 \pm 17$); however, the optically thick phase duration (OPTd) is larger than the average value found by \citet{anderson14a}. Unfortunately, as OPTd is a combination of various SN~II properties such as the hydrogen mass, radius, and interaction between ejecta and CSM, we cannot obtain direct physical constraints on the SN~2016esw progenitor from this parameter. However, it is possible from the rise, as we discuss in Section \ref{txt:rise_time}. 

Note also that the two decline rates are lower than the average found by \citet{anderson14a}, making SN~2016esw a slow-declining SN~II. It is worth noting that using the magnitude decline 50 days after maximum light proposed by \citet{faran14b}, our object (with a decline of $\sim 0.9$) mag is among the fast-declining SNe~II.

\subsubsection{Rise Time}\label{txt:rise_time}

In the standard SN~II light-curve evolution picture, after the first electromagnetic radiation emitted by the SN when the shock wave emerges from the stellar surface \citep{colgate74,falk77,klein78}, the early SN light curve is dominated by radiation from the expanding and cooling SN ejecta. The study of this early rise (i.e., the luminosity and the temperature evolution) can be used to constrain the progenitor radius \citep{waxman07,nakar10,rabinak11}.

\begin{figure}
\begin{center}
\includegraphics[width=\columnwidth]{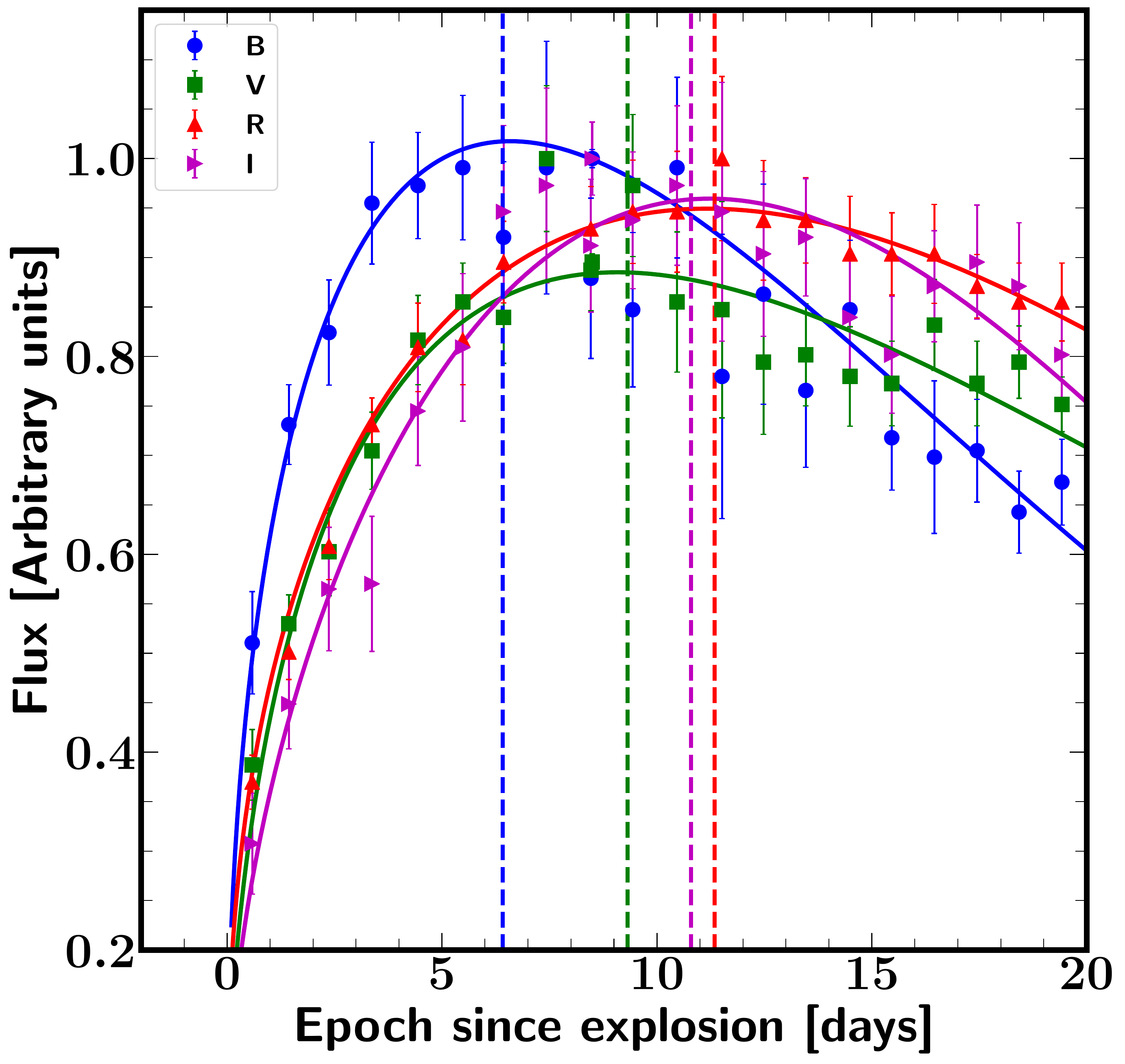}
\caption{The flux normalized to the maximum brightness of each filter ($B$, $V$, $R$, and $I$) is shown with blue dots, green squares, red up triangles, and magenta right triangles, respectively. Solid lines are the fits to Eq. (1) and vertical dotted lines mark the end of the rise.}
\label{fig:rise_fit}
\end{center}
\end{figure}

First, to derive the rise times in each band, we assume that the flux is proportional to that produced by a blackbody at a fixed wavelength \citep{waxman07,cowen10}. We fit the early-time light curves using the model defined by \citet{cowen10} and also used by \citet{roy11} and \citet{yuan16}:

\begin{equation}
f\left(t\right)~$=$~\frac{A}{e^{B\left(t-t_{0}\right)^{\alpha}}-1}\left(t-t_{0}\right)^{\beta},
\label{eq:rise}
\end{equation}

\noindent
where $t_{0}$ is the time of explosion, and $\alpha$, $\beta$, $A$, and $B$ are free parameters. Note that $\alpha$ and $\beta$ depend on the temperature evolution and the expansion of the photospheric radius, respectively. The fits for the four bands ($B$, $V$, $R$, and $I$) are shown in Figure \ref{fig:rise_fit}. Finally, the rise time is taken as the period between the explosion date (JD = 2,457,608.33 $\pm$ 0.15) and the epoch of maximum flux obtained from the derivative of Eq. \ref{eq:rise}. We derive rise-time values of 6.4, 9.6, 11.3, and 10.8 days for $B$, $V$, $R$, and $I$, respectively. These values are consistent with those derived by studies using large SN~II samples, such as \citet{gonzalezgaitan14} or \citet{gall15}. Note that these rise times are faster than the analytical and hydrodynamical rises obtained from standard RSG models \citep{gonzalezgaitan14,gall15}.

\subsubsection{Pre-explosion Radius}

To first order, the rise-time duration depends only on the progenitor radius, as the effects of the mass and energy are negligible \citep{nakar10,rabinak11,gall15,gonzalezgaitan14}. For this reason, using \citet{nakar10} analytical models, here we perform only a simple parameter study to estimate the progenitor radius. For this, we create a grid of models by varying the progenitor radius (50--700 ${\rm R}_{\odot}$ in steps of 10 ${\rm R}_{\odot}$) with a fixed explosion energy of $1.0 \times 10^{51}$ erg and a fixed stellar mass of 15 ${\rm M}_{\odot}$. These stellar mass and energy values are typical of core-collapse SNe (see \citealt{smartt09a} for a review, and references therein). 

For all these models, we derive a blackbody spectral energy distribution (SED) using the temperature defined by \citet{nakar10} in their Eq. (31) and normalised by the bolometric luminosity defined in Eq. (29). Finally, for each band, we compare the rise times from the different models (taken as the maximum) and those derived in the previous subsection. We obtain an average radius of $190 \pm 50$ ${\rm R}_{\odot}$ ($R \approx 220$, 230, 200, and 100 ${\rm R}_{\odot}$ for $B$, $V$, $R$, and $I$, respectively). Again, these values are consistent with those found in the literature. From the observed rise times, \citet{gonzalezgaitan14} and \citet{gall15} derived RSG radii of 200--500 ${\rm R}_{\odot}$ --- that is, smaller than the typical radius used in analytical or hydrodynamical RSG models ($>500$ ${\rm R}_{\odot}$). Note also that this method of extracting a rise time from the analytical models and comparing it to the observed rise time has been recently questioned by \citet{rubin16b}, as the models are valid only for a brief period (4 days). Finally, the fast rise observed in SN~II light curves could be explained by the presence of CSM rather than small RSG radii \citep{morozova17b}.

\subsubsection{Colours}

\begin{figure}
\includegraphics[width=8.5cm,height=10cm]{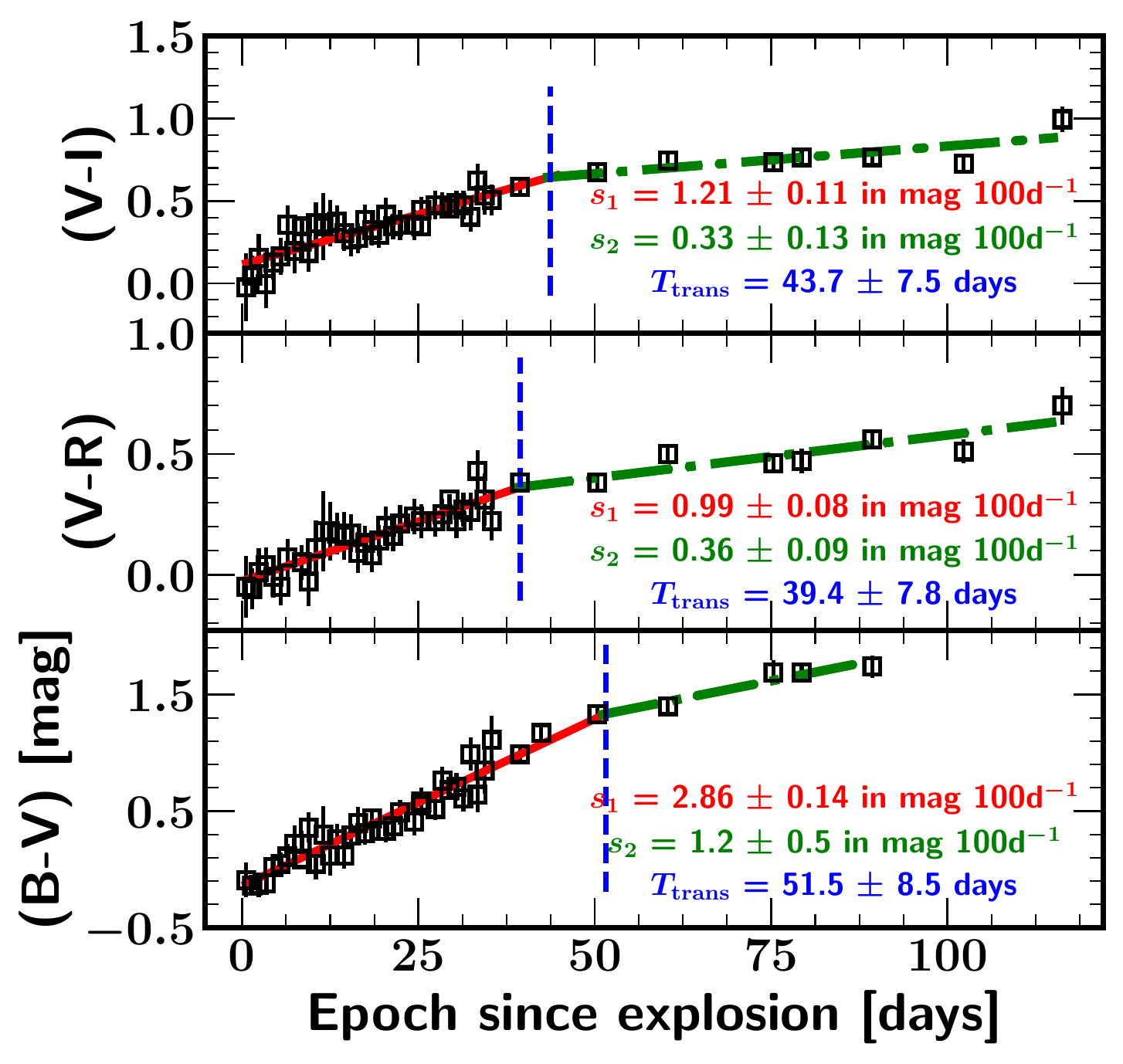}
\caption{Observed $(B-V)$, $(V-R)$, and $(V-I)$ colour evolution of SN 2016esw corrected for Milky Way extinction. In each panel, we fit the colour curves using two slopes; the first is represented by a red line while the second by a dashed-dot green line.}
\label{fig:sn2016esw_colours}
\end{figure}

The $(B-V)$, $(V-R)$, and $(V-I)$ colour curves of SN~2016esw are presented in Figure \ref{fig:sn2016esw_colours}. All the colours show the same generic behaviour expected for SNe~II \citep{dejaeger18a}: they consist of two linear regimes which are correlated. During the first $\sim 40$--50 days, the object quickly becomes redder as the SN envelope expands and cools. After this first phase, the SN colour changes more slowly as the rate of cooling decreases.

Initially, the $(B-V)$ colour increases with a rate of $2.86 \pm 0.14$ mag (100 days)$^{-1}$, followed by a slower increase of only $1.20 \pm 0.50$ mag (100 days)$^{-1}$. The transition between the two phases occurs $51.5 \pm 8.5$ days after the explosion. These values differ from those of de Jaeger et al. (in prep.), except the first slope. Indeed, with a sample of $\sim 60$ SNe~II, we derive average values for the two slopes of $s_{1,(B-V)} = 2.63 \pm 0.62$ mag (100 days)$^{-1}$ and $s_{2,(B-V)} = 0.77 \pm 0.26$ mag (100 days)$^{-1}$, and a transition time $T_{{\rm trans},(B-V)} = 37.7 \pm 4.3$ days.

Regarding $(V-R)$, the slopes are less steep, because the spectral energy distribution (SED) is less sensitive to temperature changes in the red than in the blue part of the SED. The $(V-R)$ colour curve is characterised by an initial slope of $0.99 \pm 0.08$ mag (100 days)$^{-1}$, followed by a second slope of $0.36 \pm 0.09$ mag (100 days)$^{-1}$. The transition between the two regimes occurs $39.4 \pm 7.8$ days after the explosion. 

Finally, via the same procedure, we find that the $(V-I)$ colour becomes redder with an initial slope of $1.21 \pm 0.11$ mag (100 days)$^{-1}$, followed by a slope of $0.33 \pm 0.13$ mag (100 days)$^{-1}$ and a transition at $43.7 \pm 7.5$ days. A detailed comparison with other SN~II colours is given in Section \ref{txt:compa_photometry}.

\subsubsection{Extinction and Absolute Magnitude}\label{txt:avh}

To derive intrinsic properties of the SN explosion, it is important to correct the photometry for dust reddening from the MWG and also from the host galaxy.

The Galactic extinction is straightforward using the dust map of \citet{schlafly11}. For SN~2016esw, the MWG extinction in the $V$ band is 0.188 mag\footnote{value taken from the NASA Extragalactic Database: \url{http://ned.ipac.caltech.edu/}}. Then, assuming a host-galaxy $z = 0.02825$ and a $\Lambda$CDM cosmology ($\Omega_{\rm{\Lambda}}=0.70$, $\Omega_{\rm{M}}=0.30$), with H$_{0} = 70$ km s$^{-1}$ Mpc$^{-1}$, we obtain a distance of 123.6 Mpc. This distance implies an absolute magnitude at maximum corrected only for Galactic extinction of $M_V = -17.79 \pm 0.06$ mag.

To estimate the host-galaxy extinction, different methods are used in the literature, such as the \ion{Na}{i}~D equivalent width \citep{tur03,poznanski11} or the colour excess at the end of the plateau \citep{hamuy03a,olivares10}. However, the validity of the \ion{Na}{i}~D method with low-resolution spectra (as in our case) has been questioned \citep{poznanski11,phillips13,galbany17}. The colour method has also been debated, as to use the method the authors assume the same intrinsic colour for all SNe~II, which seems to not be true (\citealt{faran14a}, de Jaeger et al. in prep.).

Despite this warning, and to compare with other studies in the literature which use one of these methods, we try to estimate the host-galaxy reddening using both methods. First, with the \ion{Na}{i}~D equivalent width (EW) method and using the relation found by \citet{tur03} ($E(B-V) \approx 0.16$ EW \ion{Na}{i}~D $-$ 0.01), we obtain $E_{\rm host}(B-V) = 0.185 \pm 0.031$ mag (EW \ion{Na}{i}~D $= 1.08 \pm 0.13$ \AA). Second, as our data do not show the transition between the plateau and the radiative tail, we cannot use the method proposed by \citet{olivares10}. Nor can we use the method derived by \citet{hamuy03a}, which consists of comparing the $(B-V)$ colour at the end of the plateau with the prototypical SN~II SN~1999em for which a precise host-galaxy extinction is well known ($E_{\rm host}(B-V) = 0.07$ mag; \citealt{leonard02}). Indeed, our $B$-band photometry stops before reaching the end of the plateau ($\sim 110$ days).

In the rest of this paper, we assume a host-galaxy extinction of $E_{\rm host}(B-V) = 0.185 \pm 0.031$ mag. Using this value, an absolute magnitude at maximum of $M_V = -18.36 \pm 0.11$ mag is derived, where the error includes the uncertainties in the magnitude and host-galaxy extinction measurements. For completeness, as it is discussed in Section \ref{txt:ifu_analyse}, we have also direct host-galaxy estimates from the IFU data, and both values are in perfect agreement ($E_{\rm host}(B-V) \approx 0.18$ mag from the IFU).

\subsection{Spectroscopy}

In this part, we describe the spectral evolution of our object, together with its velocity evolution. A full comparison with objects from the literature will be given in Section \ref{txt:discussion}.

\subsubsection{Spectral Evolution}\label{txt:spec_evolution}

Figure \ref{fig:spectral_sequence} shows the optical-wavelength spectral evolution of SN~2016esw from 0.6 to 118.5 days after explosion. The first spectrum reveals a blue featureless continuum consistent with a young core-collapse SN. The next two spectra also show a nearly featureless blue continuum, as well as a few narrow emission lines from the host galaxy (see Section \ref{txt:early_spec}). As the photosphere cools down, the broad Balmer lines grow stronger, and at day +19.5, the H$\alpha$ feature shows a boxy shape with a flat absorption profile suggesting possible ejecta-CSM interaction \citep{inserra11}. At day +26.5, the evidence of interaction start to cease, and then the object follows a typical SN~II evolution. The \ion{Fe}{II} lines start to appear in the blue part of the spectra (e.g., \ion{Fe}{II} $\lambda 5018$, \ion{Fe}{II} $\lambda 5169$), as well as the \ion{Ca}{II} near-IR triplet ($\lambda\lambda$8498, 8542, 8662) and the \ion{O}{I} $\lambda 7774$ absorption line. Present until days $+$88, these lines get stronger with time. Around days $+$33.5 and $+$47.5, close to the previous \ion{He}{I} $\lambda 5876$ line, a new feature arises and evolves with time to a strong P-Cygni profile. This line is the sodium doublet \ion{Na}{I}D $\lambda\lambda$5889, 5095. At day $+$47.5 we start to see, in addition of all the previously mentioned lines, the \ion{Sc}{II} $\lambda 5663$ and \ion{Ba}{II} $\lambda 6142$ lines. Unfortunately, the last spectrum, at day $+$118.5, is too noisy to see lines other than the prominent Balmer and \ion{Ca}{II} near-IR triplet features. It is also important to note that between epochs $+$19.5 and $+$33.5 days after the explosion, the H$\alpha$ P-Cygni profile shows an atypical shape produced by ``Cachito'', which is an extra absorption component on the blue side of H$\alpha$ associated with \ion{Si}{II} at early epochs \citep{gutierrez17a}. When it disappears the P-Cygni profile becomes normal.

\subsubsection{Early-time spectroscopy}\label{txt:early_spec}

Recent studies \citep{galyam14,shivvers15,khazov16,yaron17} have shown that early-time spectra (1--2 days after the explosion) can be dominated by strong, high-ionisation emission lines produced by the recombination of dense CSM ionised by first the shock-breakout flash \citep{galyam14} and second by ultraviolet light emitted during the week after the RSG explosion \citep{dessart17}. Early-time spectral observations are challenging and still uncommon in the literature; thus, here we present a comparison with SN~2013fs (iPTF13dqy; \citealt{yaron17}).

\begin{figure*}
\includegraphics[width=\textwidth]{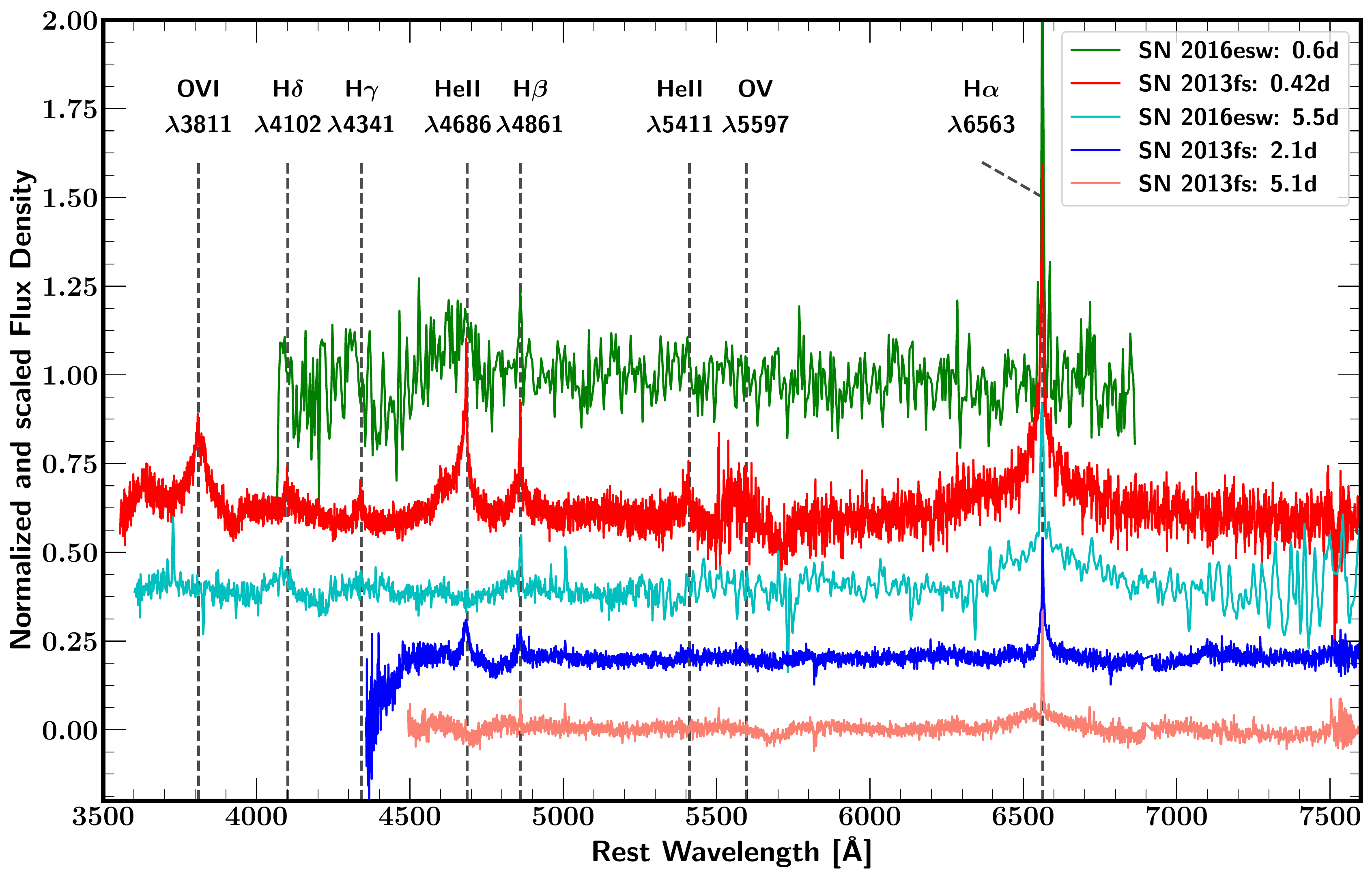}
\caption{Early-time spectroscopy of SN~2013fs \citep{yaron17} and SN~2016esw. All of the spectra have been normalised and scaled for clarity.}
\label{fig:early_spec}
\end{figure*}

In Figure \ref{fig:early_spec}, we compare our two earliest spectra (0.6\,d and 5.5\,d\footnote{For clarity, we omit the spectrum taken 6.5\,d after the explosion; it is nearly identical to the one taken at 5.5\,d.}) of SN~2016esw with those of SN~2013fs at epoch of 0.42\,d (10.1\,hr), 2.1\,d, and 5.1\,d. Unlike SN~2013fs, our first spectrum does not exhibit strong, narrow emission lines of \ion{He}{ii} $\lambda$4686, \ion{He}{ii} $\lambda$5411, or \ion{O}{V} $\lambda$5597. The absence of high-ionisation oxygen lines is expected, as they should disappear within $\sim 11$\,hr after the explosion \citep{yaron17}. However, the \ion{He}{ii} emission lines from a strong progenitor wind should be visible until $\sim 5$\,d after the explosion \citep{khazov16}. Our two spectra taken at 5.5\,d and 6.5\,d do not reveal \ion{He}{ii} emission lines, but rather a blue continuum with weak H$\alpha$ emission. These two spectra are very similar to those of SN~2013fs at 5.1\,d after explosion. At this epoch, the \ion{He}{ii} emission lines seen at earlier epochs in spectra of SN~2013fs disappear completely; instead, there is an almost featureless blue continuum (with only weak Balmer lines).

In SN~2016esw, the absence of high-ionisation emission lines at early phases could suggest a lack of dense CSM around the progenitor; alternatively, perhaps it was not observed at sufficiently early epochs \citep{khazov16,dessart17}. As our first observed spectrum was taken only 0.6\,d after the explosion, the most plausible explanation is the absence of dense material near the star's surface. Given that we detect signs of interaction at later epoch (19.5\,d after the explosion; see Section \ref{txt:spec_evolution}), we believe that the progenitor was surrounded by low-density CSM, but not near the star's surface and not ejected during the final stage ($\sim 1$\,yr) prior to explosion, in contrast to the case of SN~2013fs.

\begin{figure}
\includegraphics[width=\columnwidth]{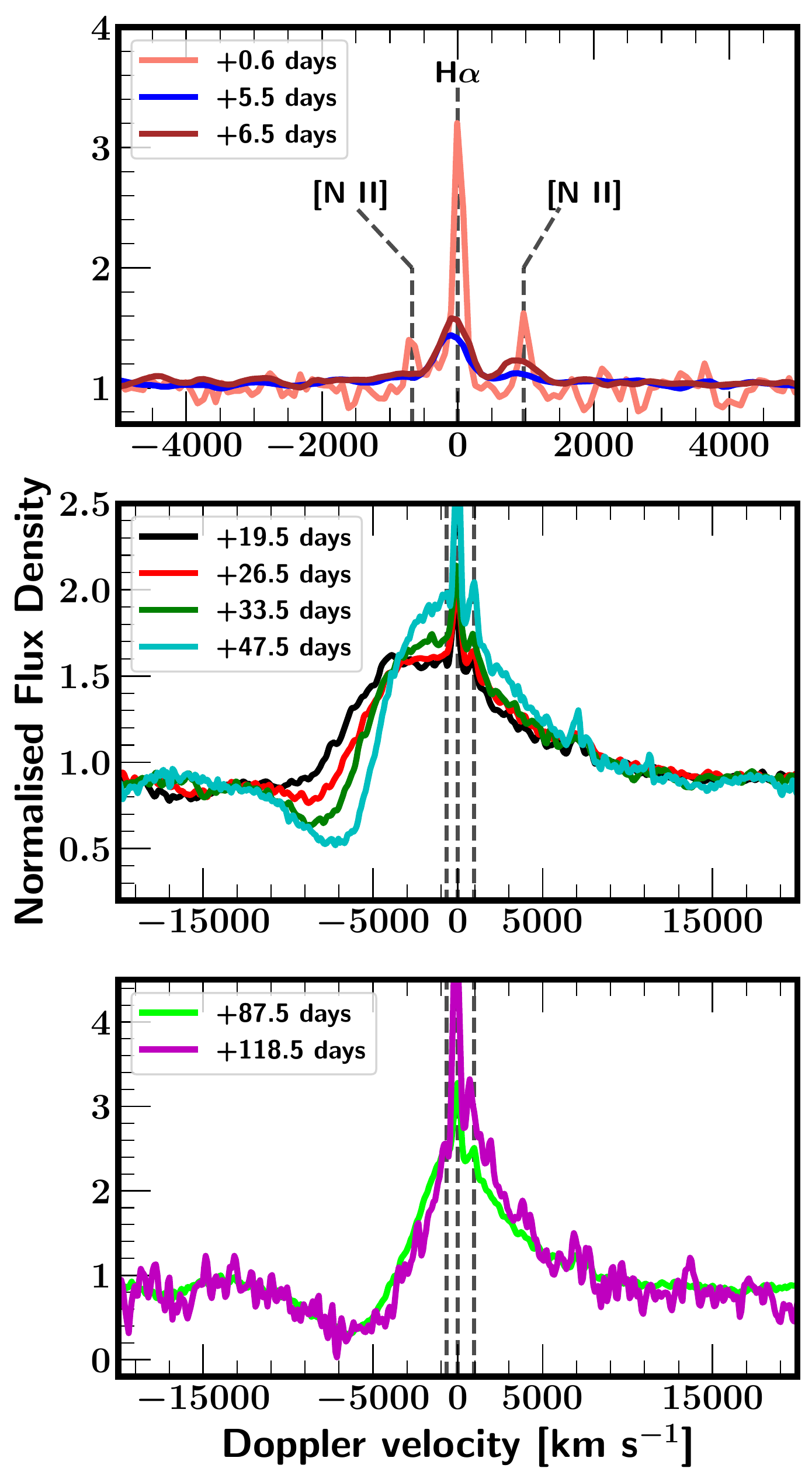}
\caption{Close-up views centred on H$\alpha$. The spectra are separated in three phases: early-time (top), plateau (middle), and plateau end (bottom). In all the panels, the vertical dashed lines correspond to [N~II] $\lambda\lambda$6548, 6583 and H$\alpha$. Narrow (spectrally unresolved) emission lines are almost certainly produced by a contaminating H~II region.}
\label{fig:Halpha_zoom}
\end{figure}

Finally, all three of our early-time spectra exhibit very narrow H$\alpha$ (and H$\beta$) emission lines which could be caused by a contaminating H~II region or by CSM interaction. In Figure \ref{fig:Halpha_zoom}, we illustrate a close-up view centred on H$\alpha$. The spectra show characteristic lines of an H~II region (e.g., [N~II] $\lambda\lambda$6548, 6583); thus, we believe that the spectrally unresolved H$\alpha$ line superposed on the broader H$\alpha$ profile is produced mainly or entirely by host-galaxy contamination. Note also that this line is present in all of our spectra, even those without clear signs of interaction, supporting a host-galaxy origin rather than CSM interaction. 

\subsubsection{Expansion Velocity and Temperature}

\begin{figure}
\includegraphics[width=\columnwidth]{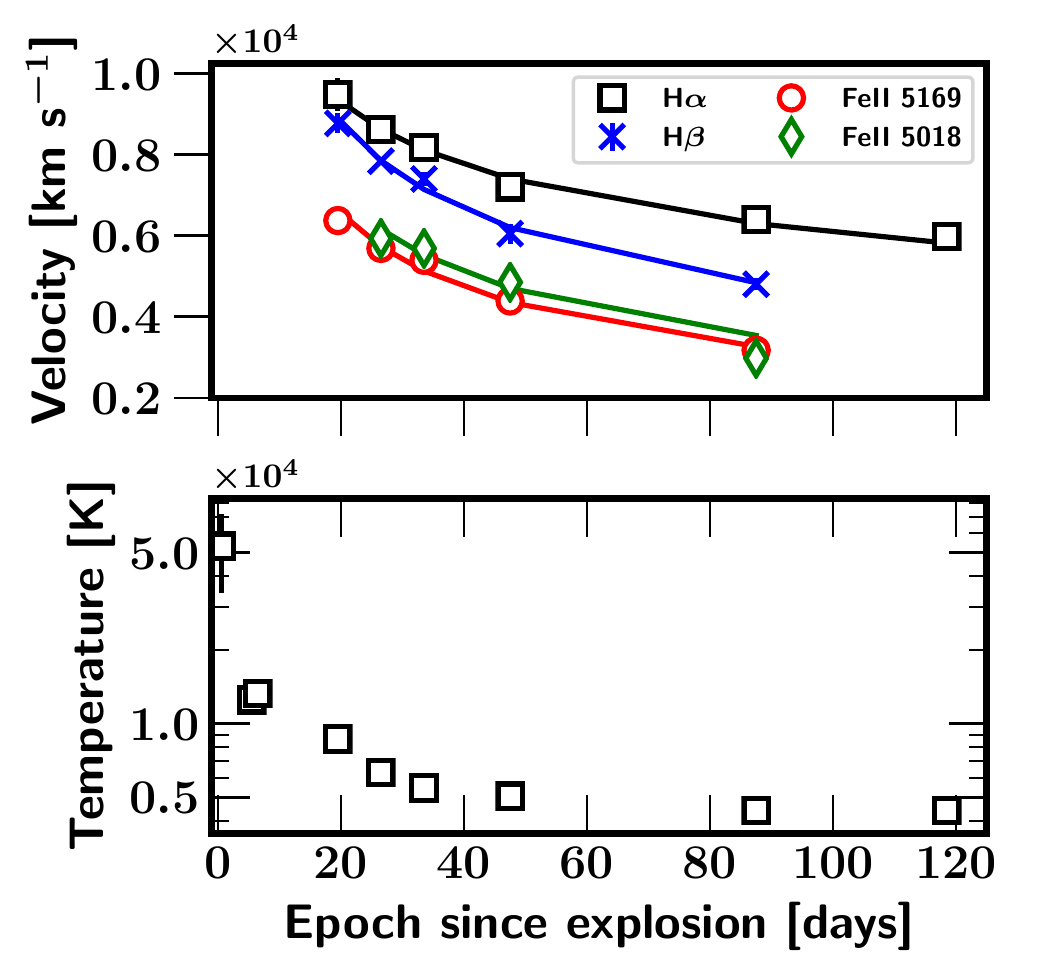}
\caption{\textit{Top:} Expansion velocities of H$\alpha$, H$\beta$, \ion{Fe}{ii} $\lambda$5018, and \ion{Fe}{ii} $\lambda$5169. \textit{Bottom:} Blackbody temperature evolution.}
\label{fig:vel_sn2016esw}
\end{figure}

We investigate the expansion velocities of H$\alpha$ $\lambda$6563, together with lines used for the standard candle method \citep{hamuy02,nugent06,andrea10,poznanski10,dejaeger17a,dejaeger17b} --- H$\beta$ $\lambda$4861, \ion{Fe}{ii}$\lambda$ 5018, and \ion{Fe}{ii} $\lambda$5169. All of the velocities are measured through the minimum flux of the absorption component of the P-Cygni line profile after correcting the spectra for the heliocentric redshift of the host galaxies. Uncertainties in the velocities were obtained by measuring many times the minimum of the absorption (changing the continuum fit each time) and determining their standard deviation. We are not able to measure the velocities in the first three spectra, as the Balmer lines are strongly contaminated by the host-galaxy emission lines and no clear absorption is seen. All of the velocities are reported in Table \ref{tab:log_velocity} and plotted in Figure \ref{fig:vel_sn2016esw}. 

SN~II velocity evolution is well known and decreases following a power law \citep{hamuyphd}: with time, the photosphere advances deeper into the ejecta, and thus we see material moving at progressively lower velocities. The SN~2016esw velocity evolution follows this behaviour and is characterised by power-law exponents of $-0.264 \pm 0.011$, $-0.405 \pm 0.014$, $-0.459 \pm 0.020$, and $-0.465 \pm 0.018$ for H$\alpha$ $\lambda$6563, H$\beta$ $\lambda$4861, \ion{Fe}{ii} $\lambda$5018, and \ion{Fe}{ii} $\lambda$5169, respectively (fits are shown in \ref{fig:vel_sn2016esw}). These values are very consistent with those found in the literature. \citet{dejaeger17a}, using the whole Carnegie Supernova Project I sample, derived for the H$\beta$ velocity an exponent of $-0.407 \pm 0.173$. On the other hand, using the \ion{Fe}{ii} $\lambda$5169 line, \citet{nugent06} and \citet{dejaeger15b} derived values of $-0.464 \pm 0.017$ and $-0.55 \pm 0.20$, respectively. Finally, as expected, since the H$\alpha$ and H$\beta$ lines are formed above the photosphere (i.e., at larger radii), their velocities are higher than those derived for the \ion{Fe}{ii} lines which are better connected to the photospheric velocity.

For all the spectra, we perform a blackbody fit and derive the temperature, as shown in Table \ref{tab:log_velocity} and in Figure \ref{fig:vel_sn2016esw}. The temperature evolution is rather normal for a SN~II, with a broken power law described by a rapid decline at early phases followed by a slower decline at later epochs \citep{faran17}. Indeed, the early spectra taken only 5.5--6.5 days after the explosion show a blue continuum with a hot blackbody temperature of $T_{\rm bb} \approx 12,500$ K. Subsequently the temperature drops about 5000 K in only $\sim 30$ days. Finally, after this first decline ($\sim 30$ days after the explosion), the hydrogen envelope starts to recombine and the temperature evolves more slowly, with a decrease of only $\sim 1000$ K in $\sim 50$ days.\\

\begin{table*}
\begin{minipage}{126mm}
	\centering
	\caption{Observed Blackbody Temperatures and Expansion Velocities of SN~2016esw.}
\begin{tabular}{lcccccr}
		\hline
		UT Date & Epoch & $T_{\rm bb}$ & $v$(H$\alpha$) & $v$(H$\beta$) & $v$(\ion{Fe}{II} $\lambda 5018$) & $v$(\ion{Fe}{II} $\lambda 5169$) \\
		 &(days) & (K) &(km s$^{-1}$) & (km s$^{-1}$) & (km s$^{-1}$) &(km s$^{-1}$)\\
		\hline
		2016-08-08.41 &0.6 & 53,150 (19,100) & $\cdots$ ($\cdots$) & $\cdots$ ($\cdots$) & $\cdots$ ($\cdots$) &$\cdots$ ($\cdots$)\\
		2016-08-13.28 &5.5 & 12,500 (1,100) & $\cdots$ ($\cdots$) & $\cdots$ ($\cdots$) & $\cdots$ ($\cdots$) &$\cdots$ ($\cdots$)\\
		2016-08-14.29 &6.5 & 13,330 (1,200) & $\cdots$ ($\cdots$) & $\cdots$ ($\cdots$) & $\cdots$ ($\cdots$) &$\cdots$ ($\cdots$)\\
		2016-08-27.22 &19.5 & 8650 (500) & 9480 (410) & 8770 (250) &$\cdots$ ($\cdots$) & 6375 (270)\\
		2016-09-03.22 &26.5 & 6300 (350) & 8615 (100) & 7840 (100) & 5930 (120) & 5690 (150)\\
		2016-09-10.20 &33.5 & 5450 (300) & 8175 (80) & 7400 (180) & 5690 (110) & 5390 (170)\\
		2016-09-24.16 &47.5 & 5050 (300) & 7200 (130) & 6050 (250) & 4850 (100) & 4390 (180)\\
		2016-11-03.16 &87.5 & 4420 (250) & 6400 (180) & 4800 (150) & 2980 (200) & 3175 (130)\\
		2016-12-04.13 &118.5 & 4410 (300) & 5985 (300) & $\cdots$ ($\cdots$) & $\cdots$ ($\cdots$) & $\cdots$ ($\cdots$)\\
		\hline
\end{tabular}
\label{tab:log_velocity}
\end{minipage}
\end{table*}

\section{Discussion}\label{txt:discussion}

Here, we compare our object properties with those of a sample of other well-observed SNe~II, including the well-studied prototypical SN~II SN~1999em ($M_V \approx -16.94$ mag; \citealt{hamuy01,leonard02,elmhamdi03}). The sample also contains the prototype of fast-declining SNe~II (historically SNe~IIL), SN~1979C ($M_V \approx -19.45$ mag; \citealt{bra81,devaucouleurs81}), together with well-observed SNe~II similar to SN~2016esw in optically thick duration (SN~2004et; $M_V \approx -17.5$ mag; \citealt{sahu06,maguire10b}) or brightness: SN~2007od ($M_V \approx -18.0$ mag; \citealt{inserra11}), SN~2013by ($M_V \approx -18.2$ mag; \citealt{valenti15}), and SN~2013ej ($M_V \approx -17.8$ mag; \citealt{valenti14,bose15,huang15,mauerhan16,dhungana16}). We also select two SNe~II from the moderately luminous SN~II sample of \citet{inserra13}, as both objects show ejecta-CSM interaction: SN~2007pk ($M_V \approx -18.44$ mag) and SN~2009dd ($M_V \approx -17.46$ mag). A subluminous SN~II is also added: SN~2005cs ($M_R \approx -15.2$ mag; \citealt{pastorello09}). The explosion epochs (second column), distances\footnote{Hubble constant of 70 km s$^{-1}$ Mpc$^{-1}$ and a $\Lambda$CDM Universe with $\Omega_{\rm{M}}=0.3$ and $\Omega_{\rm{\Lambda}}=0.7$.} (third column), and extinctions (fourth column) for each SN are detailed in Table \ref{tab:sn_litterature}.\

\begin{table}
	\caption{Properties of Our Sample of Type II Supernovae.}
	\begin{tabular}{lllll} 
		\hline
		SN & Explosion & $D$ & $E(B-V)_{\rm tot}$ & Ref\\
		 & Epoch (MJD) &(Mpc) & (MWG+host; mag) & \\
		\hline
		1979C & 2,443,967 &15.2 &0.023+0.130 &1,2,3\\
		1999em & 2,451,476 &11.7 &0.030+0.070 &4\\
		2004et & 2,453,270 &5.9 &0.300+0.110 &5\\
		2005cs & 2,453,549 &7.1 &0.030+0.020 &6\\
		2007od & 2,454,404 &24.9 &0.038+0.000 &7\\
		2007pk & 2,454,412 &70.1 &0.050+0.060 &8\\
		2009dd & 2,454,925 &14.1 &0.020+0.430 &8\\
		2013by & 2,456,404 &14.8 &0.195+0.000 &9\\
		2013ej & 2,456,497 &9.6 &0.061+0.000 &10\\
		2016esw & 2,457,608 &123.6 &0.061+0.185 &11\\
		\hline
	\end{tabular}
\medskip{References: (1) \citet{weiler86}, (2) \citet{freedman01}, (3) \citet{devaucouleurs81}, (4) \citet{leonard02}, (5) \citet{maguire10b}, (6) \citet{pastorello09}, (7) \citet{inserra11}, (8) \citet{inserra13}, (9) \citet{valenti15}, (10) \citet{bose15}, (11) This work.}
\label{tab:sn_litterature}
\end{table}

\subsection{Photometry}\label{txt:compa_photometry}

\begin{figure}
\includegraphics[width=\columnwidth]{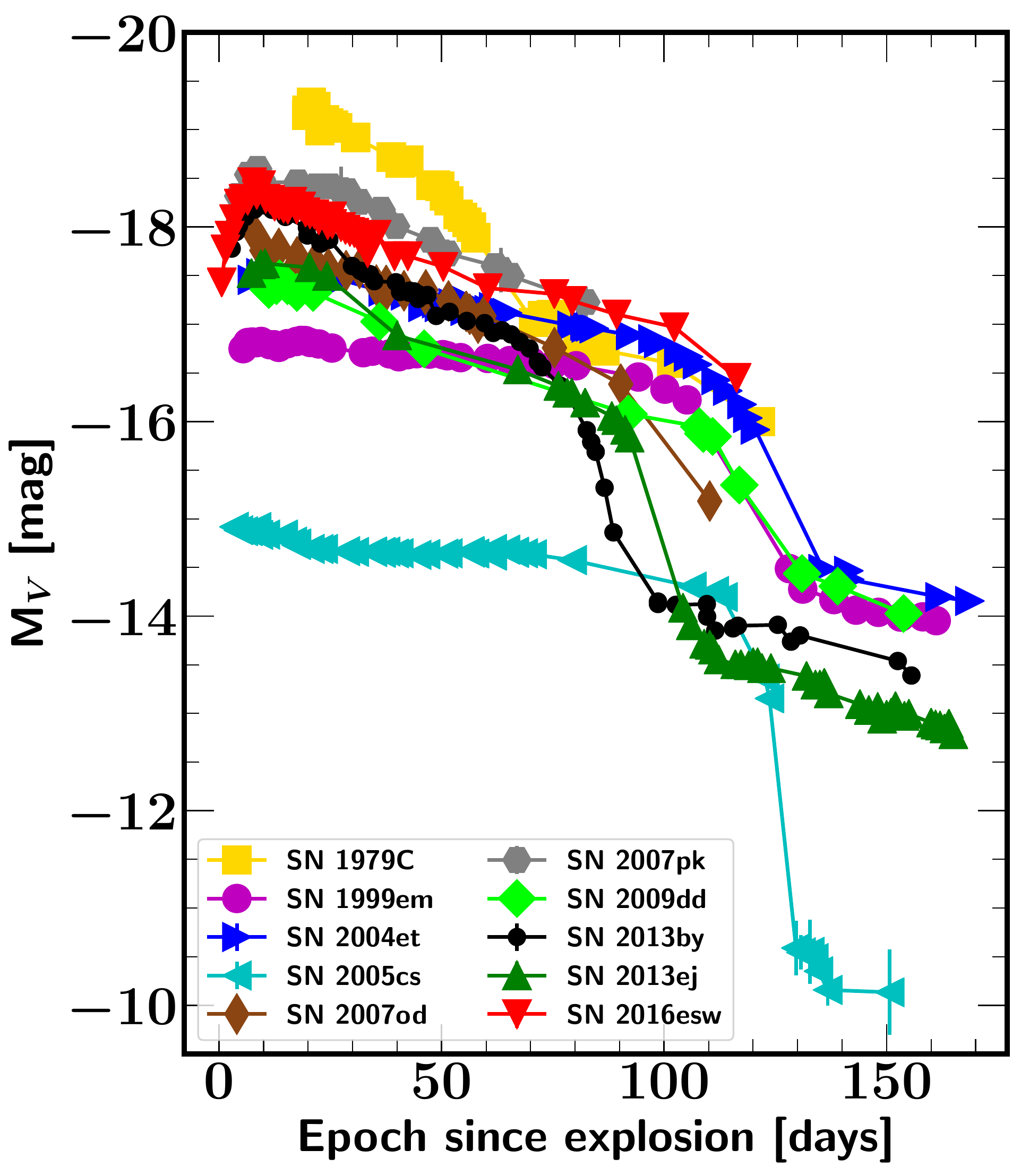}
\caption{Comparison of the $V$-band light curve of SN~2016esw with those of other well-observed SNe~II: SN~1999em \citep{leonard02}, SN~2004et \citep{maguire10b}, SN~2005cs \citet{pastorello09}, SN~2007od \citep{inserra11}, SN~2007pk \citep{inserra13}, SN~2009dd \citep{inserra13}, SN~2013by \citep{valenti15}, and SN~2013ej\citep{huang15}. The fast-declining SN~1979C \citet{devaucouleurs81} is also shown.}
\label{fig:compa_mag}
\end{figure}

Figure \ref{fig:compa_mag} shows a comparison of the $V$-band light curves of SN~2016esw and other well-observed SNe~II. We clearly see that SN~2016esw is much more luminous than the prototypical SN~II (SN~1999em);it is in the bright tail end of the SN~II luminosity distribution. For example, with a peak absolute magnitude of $-$18.32, our object is more luminous than SN~2013by ($M_V \approx -18.2$ mag) or SN~2007od ($M_V \approx -18.0$ mag). However, SN~2016esw is fainter than SN~2007pk ($M_V \approx -18.44$ mag) and SN~1979C ($M_V \approx -19.45$ mag). SN~2007pk shows strong ejecta-CSM interaction and was first classified as a SN~IIn \citep{parisky07}. On the other hand, SN~1979C has a $V$-band light curve with a very steep decline during the photospheric phase.

Regarding the $V$-band light-curve behaviour, SN~2016esw has a longer recombination phase than the majority of luminous SNe~II (SN~2007od, SN~2009dd, SN~2013by, and SN~2013ej). However, the $V$-band light curve is comparable to that of SN~2004et, with the same recombination-phase duration ($\sim 110$ days) but with a brighter plateau (by $\sim 0.2$ mag). Though the late-time light curves of SN~2016esw and SN~2004et are similar, at early times they show differences possibly caused by CSM interaction. For SN~2016esw, the initial decline after maximum brightness is much steeper than for SN~2004et. The $V$ light curve of SN~2016esw also displays similarities to that of SN~2007pk regarding the brightness, but after 40 days SN~2007pk is more linear than our object. 

\begin{figure}
\includegraphics[width=\columnwidth]{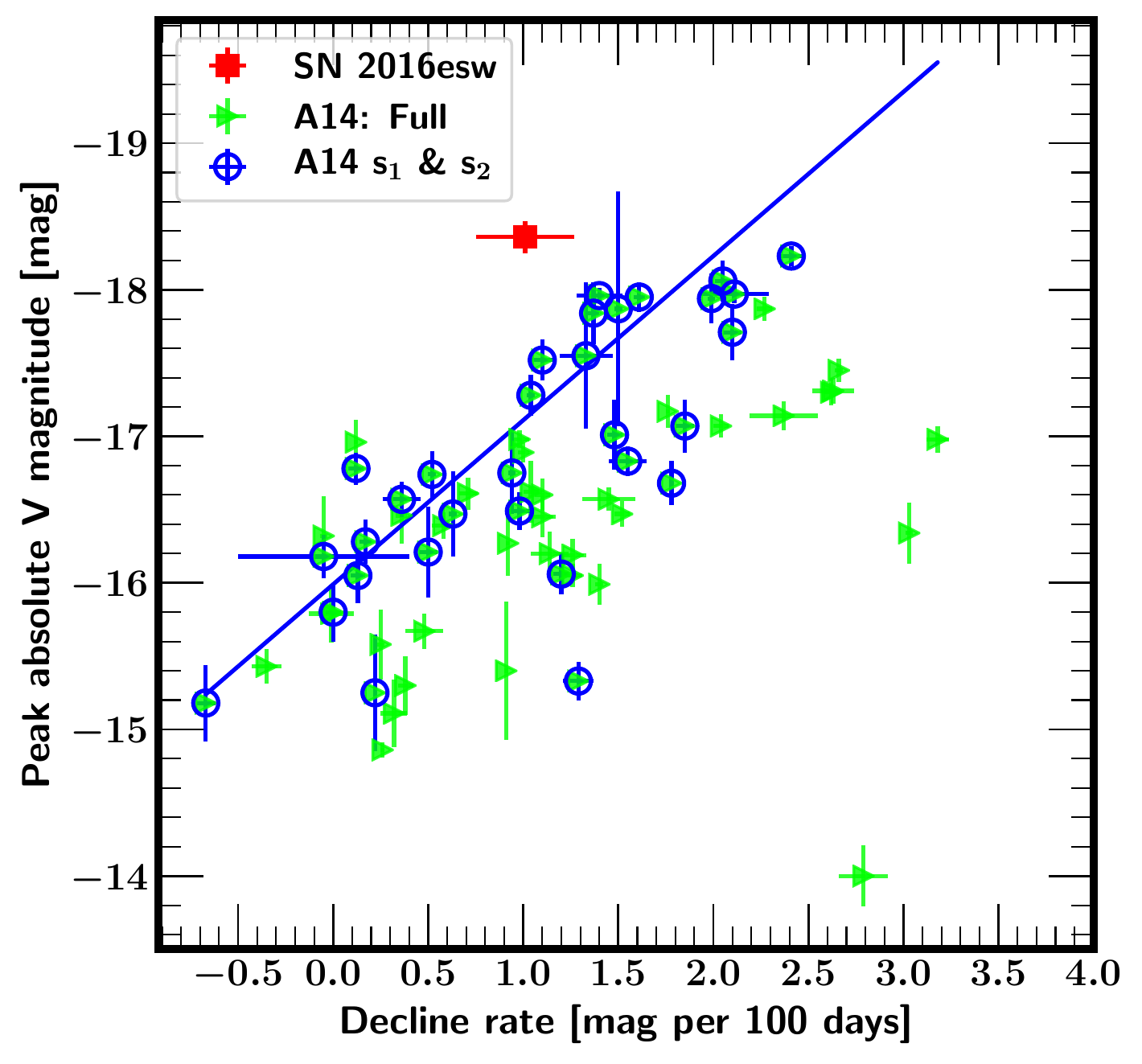}
\caption{Correlation between $M_{\rm max}$ and $s_{2}$. Green right triangles represent the full sample used by \citet{anderson14a} while the red square is SN~2016esw. The blue line shows the linear relation ($M_{\rm max} = -1.12\,s_{2} - 15.99$) between $M_{\rm max}$ and $s_{2}$ derived by \citet{anderson14a} using only SNe with $s_{1}$ and $s_{2}$ (blue circles).}
\label{fig:s2_sample}
\end{figure}

It is also important to note that even if the $V$-band light curve belongs to the SN~II family, SN~2016esw seems to not follow the correlation found by \citet{anderson14a} between the absolute peak magnitude and the slope of the plateau ($M_V = -1.12 \, s_{2} - 15.99$ mag). Even if SN~2016esw is not considered an outlier according to the \citet{chauvenet1863} criterion, it exhibits a flatter slope than the other luminous SNe~II --- that is, for its absolute peak magnitude ($M_V =-18.36$ mag), SN~2016esw should have a slope of $\sim 2.10$ mag (100 days)$^{-1}$ instead of $1.01\pm 0.26$ mag (100 days)$^{-1}$. This could be important for cosmology with SNe~II: to reduce the Hubble-diagram scatter, one of the methods used to standardise SNe~II is based on the absolute magnitude vs. $s_{2}$ correlation (photometric colour method; \citet{dejaeger15b}). Figure \ref{fig:s2_sample} shows the peak magnitude vs. $s_{2}$ correlation from \citet{anderson14a} and where our object lies.

\begin{figure}
\includegraphics[width=\columnwidth]{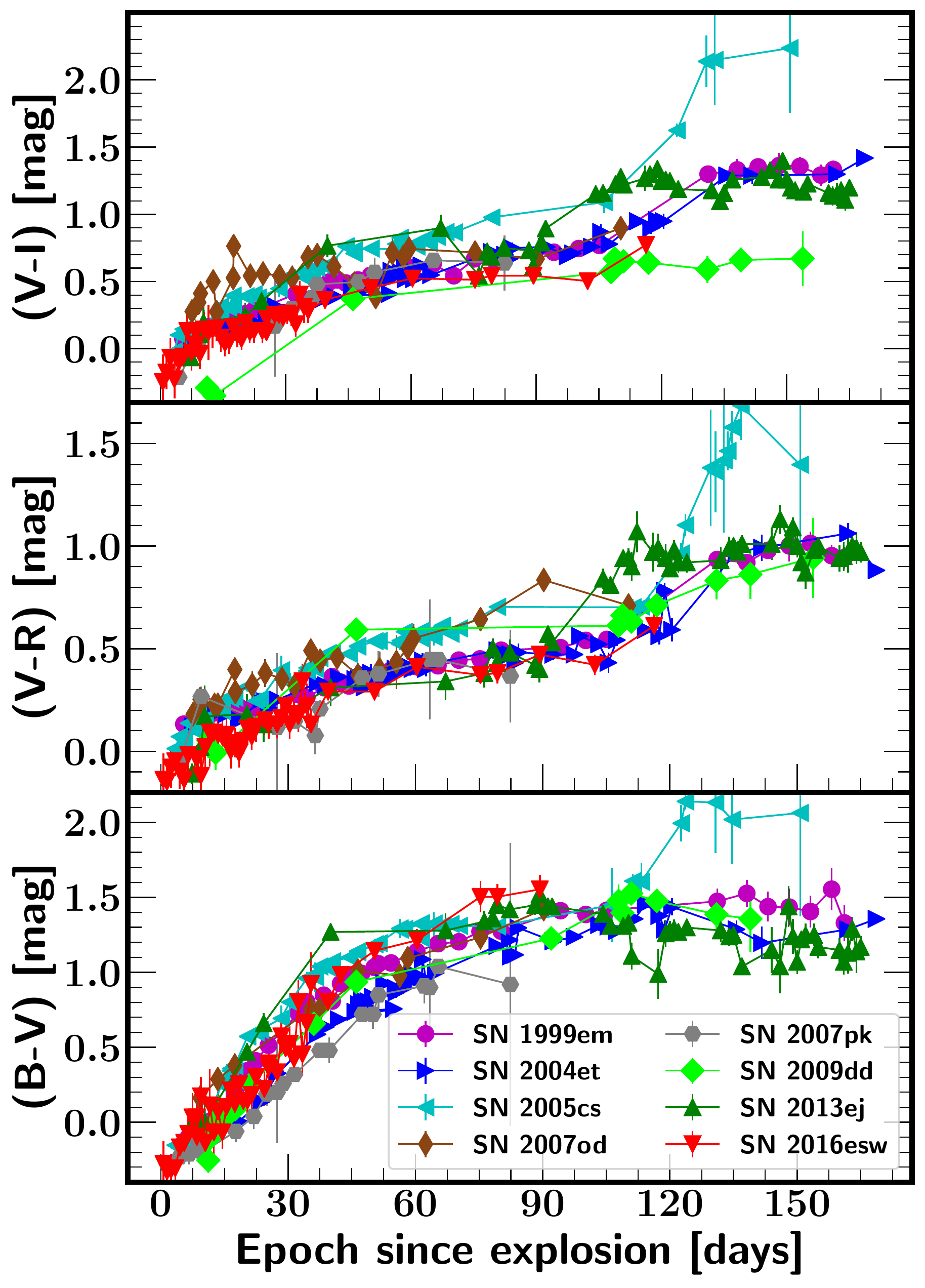}
\caption{Comparison of the $(B-V)$, $(V-R)$, and $(V-I)$ colour curves of SN~2016esw with those of other well-observed SNe~II: SN~1999em \citep{leonard02}, SN~2004et \citep{maguire10b}, SN~2005cs \citet{pastorello09}, SN~2007od \citep{inserra11}, SN~2007pk \citep{inserra13}, SN~2009dd \citep{inserra13}, and SN~2013ej\citep{huang15}.}
\label{fig:compa_colours}
\end{figure}

Regarding the colour differences, in Figure \ref{fig:compa_colours}, we show a comparison of the $(B-V)$, $(V-R)$, and $(V-I)$ colour curves with those of the well-observed SNe~II used in the previous paragraph. The $(B-V)$ trend of SN~2016esw is very similar to that of normal SNe~II (e.g., SN~1999em), and the $(V-R)$ and $(V-I)$ colour curves do not show a red peak after $\sim 100$ days as seen for SN~2005cs. Finally, the colours also confirm the similarities to SN~2004et and SN~2007pk, with nearly identical $(V-R)$ and $(V-I)$ colour evolution. However, at late epochs SN~2016esw seems to have a redder $(B-V)$ colour than do SN~2004et or SN~2007pk.

\subsection{Spectroscopy}\label{compa_spectroscopy}

\begin{figure*}
\includegraphics[width=2.0\columnwidth]{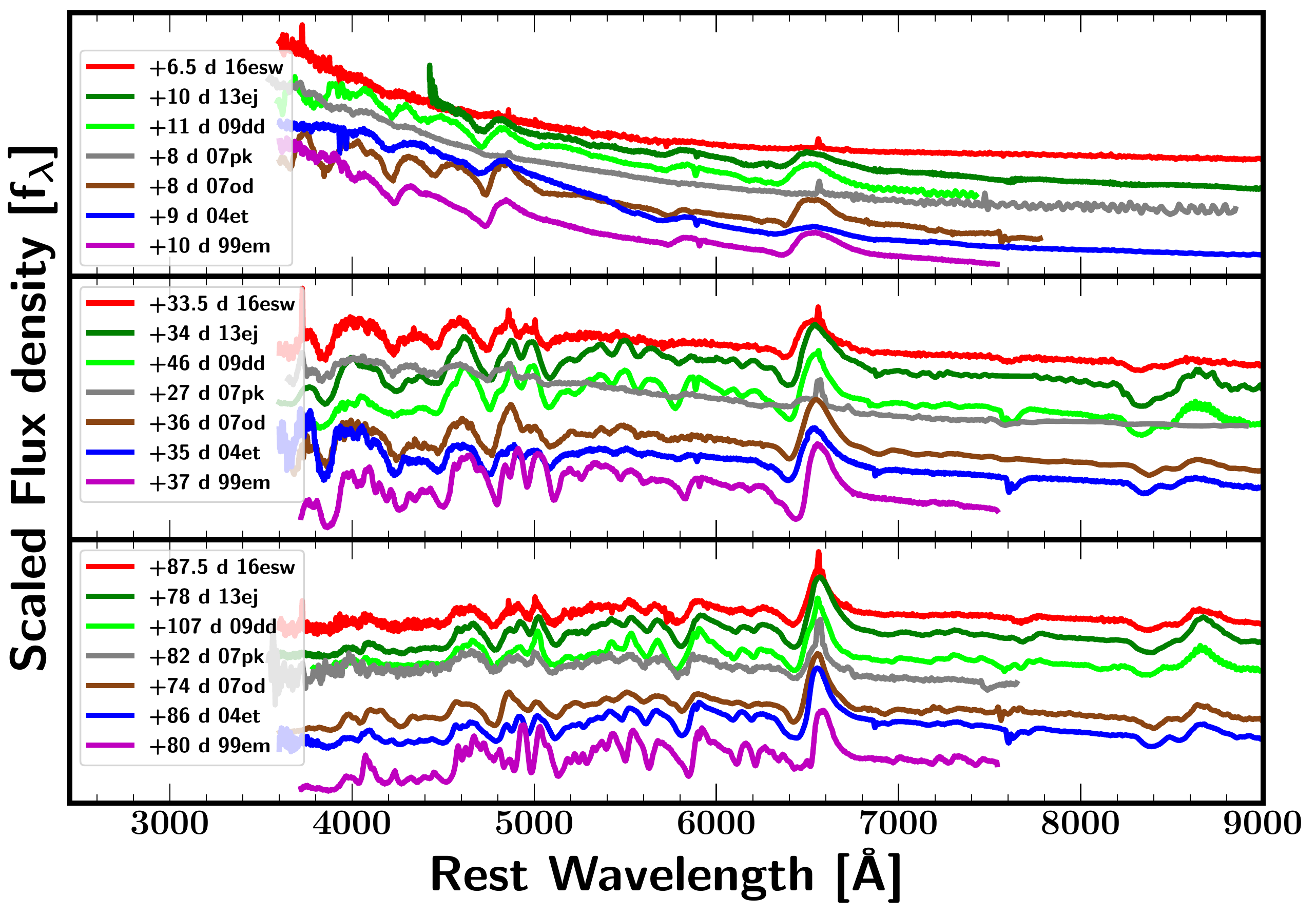}
\caption{Comparison of early-time (10 days), plateau (35 days), and end of the plateau (80 days) spectra of SN~2016esw with those of other well-studied Type II SNe: SN~1999em \citep{baron00}, SN~2004et \citep{sahu06,maguire10b,guillochon16}, SN~2007od \citep{inserra11,hicken17}, SN~2007pk \citep{inserra13,hicken17}, SN~2009dd \citep{inserra13}, and SN~2013j \citep{huang15,dhungana16}. All comparison spectra are corrected for extinction and redshift (adopted values in Table \ref{tab:sn_litterature}).}
\label{fig:compa_spectra}
\end{figure*}

Figure \ref{fig:compa_spectra} displays a spectroscopic comparison between SN~2016esw and our well-studied SN~II sample defined in Table \ref{tab:sn_litterature} at three different epochs: early phase ($\sim 10$ days), plateau phase ($\sim 35$ days), and the end of the recombination phase ($\sim 80$ days).

Qualitatively, the comparison shows that the SN~2016esw spectra are broadly similar to others after 30 days past explosion. In particular, after this epoch, SN~2016esw and SN~2004et become mostly indistinguishable in terms of observable line features and their evolution. Both of these SNe~II also exhibit identical absorption-component strengths, indicating similar temperatures. However, the spectra of SN~2016esw have shallower H$\alpha$ absorption, which is attributed to differences in progenitor properties; for example, fast-declining SNe~II have smaller absorption \citep{gutierrez14}.

At early epochs, SN~2016esw (along with SN~2007pk) is quite different from the other SNe~II, with a featureless blue continuum and resolved narrow Balmer emission lines. As SN~2007pk is known to have relatively strong ejecta-CSM interaction \citep{inserra13}, we believe that the SN~2016esw early-time spectra are also contaminated by some interaction between the ejecta and CSM; evidence for CSM is also visible in the spectrum taken 19.5\,d after the explosion (see Section \ref{txt:spec_evolution}). The fact that these two SNe~II are also among the most luminous SNe~II advocates in favour of CSM interaction at early epochs, as some luminosity could be added from the CSM interaction. Additionally, the plateau length could be supported by such interaction. Even if the effect on the plateau length is very small with respect to the impact on the early-time light curve, \citet{morozova17} demonstrated that hydrodynamical models with a dense wind fit the data better than those without the dense wind (models with CSM have a longer plateau).

Finally, at late epochs, both objects also display very similar spectra, with nearly identical absorption-component strengths (e.g., H$\alpha$). However, at epoch $\sim +30$ days after the explosion, the two spectra differ. This difference might be explained by the fact that their spectra represent snapshots of the transition from interacting SNe~II to normal SNe~II, but the phases aren't quite the same and the transitions may have proceeded at different rates.

\begin{figure}
\includegraphics[width=\columnwidth]{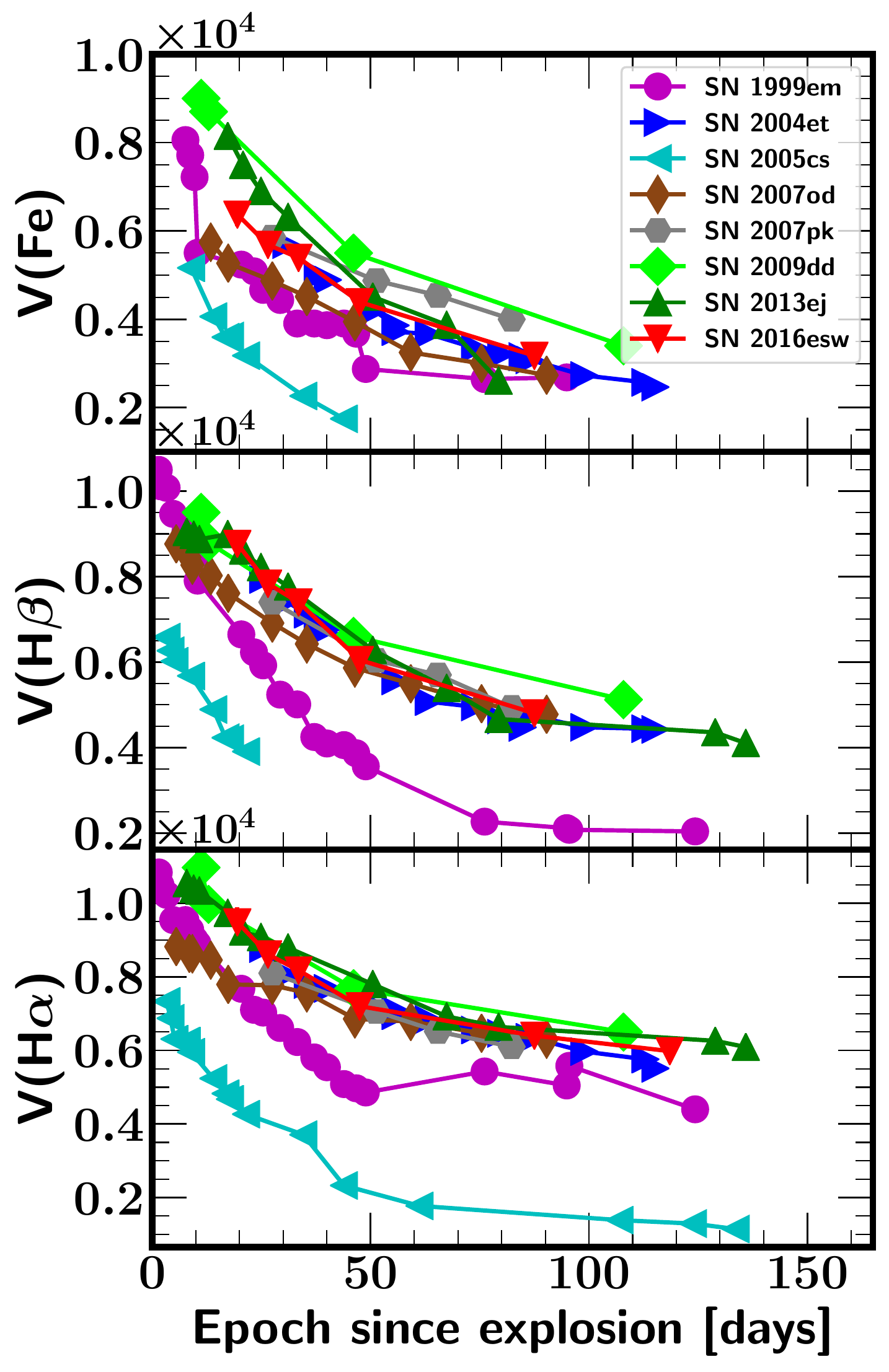}
\caption{Evolution of the H$\alpha$ (bottom), H$\beta$ (middle), and \ion{Fe}{ii} $\lambda 5169$ (top) velocities of SN~2016ew compared with those of SN~1999em \citep{leonard02}, SN~2004et \citep{maguire10b}, SN~2005cs \citet{pastorello06,pastorello09}, SN~2007od \citep{inserra11}, SN~2007pk \citep{inserra13}, SN~2009dd \citep{inserra13}, and SN~2013ej \citep{huang15}. All velocities are expressed in units of $10^4$ km s$^{-1}$.}
\label{fig:compa_vel}
\end{figure}

In Figure \ref{fig:compa_vel}, we compare the H$\alpha$, H$\beta$, and \ion{Fe}{ii} $\lambda 5169$ velocity evolution of SN~2016esw with those of our comparison sample (SN~1999em, SN~2004et, SN~2005cs, SN~2007od, SN~2007pk, SN~2009dd, and SN~2013ej). For the three lines, SN~2016esw velocities are comparable to those of the majority of our sample but higher than the prototypical Type II SN~1999em and much higher than SN~2005cs, a low-luminosity SN~II which is known to have slow photospheric expansion. It is important to note again the similarity with SN~2004et and SN~2007pk, where the velocities are almost identical to those of our object.

\subsection{Stellar Populations at the SN Location}\label{txt:ifu_analyse}

\begin{figure*}
\includegraphics[width=\textwidth]{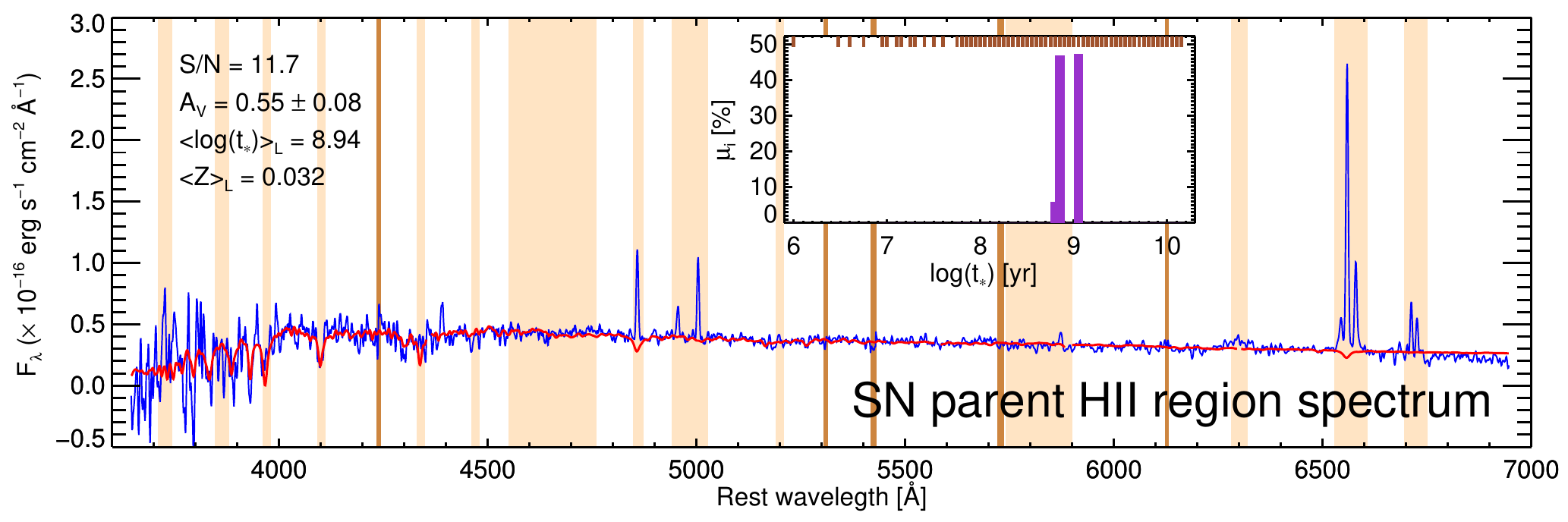}
\includegraphics[trim=-1cm 0cm -0.4cm 0cm,clip=true,width=\textwidth]{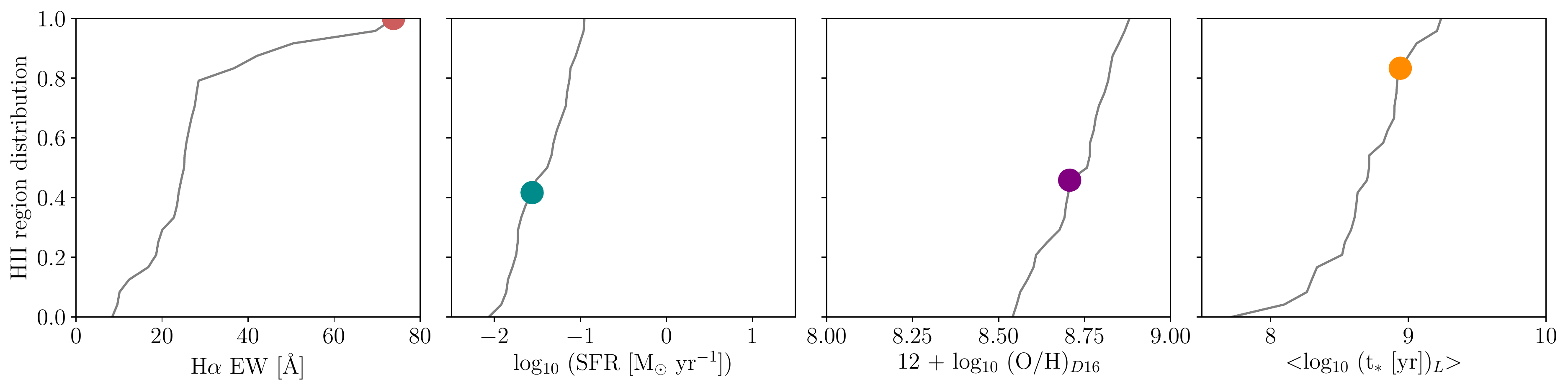}
\caption{{\it Top:} SN 2016esw parent H~II region spectrum and best simple stellar population fit from STARLIGHT (red). Vertical strips represent masked regions in the fit due to known emission lines (such as Balmer lines, oxygen, WR bumps, etc.; in beige) and regions with night-sky lines (in brown). The inner panel shows the star-formation history of the spectrum (in purple), where the upper brown tick marks represent the ages of the models used in the fit. The signal-to-noise ratio of the spectrum, optical extinction, average stellar age, and average metallicity are reported in the upper-left corners.
{\it Bottom:} Distributions of H$\alpha$ equivalent width, star-formation rate, oxygen abundance, and average stellar age of all 25 H~II regions in CGCG 229-009. Dotted colours represents the position of the SN 2016esw parent H~II region in these distributions.} 
\label{fig:ssp_local}
\end{figure*}

\begin{figure*}
\includegraphics[width=\textwidth]{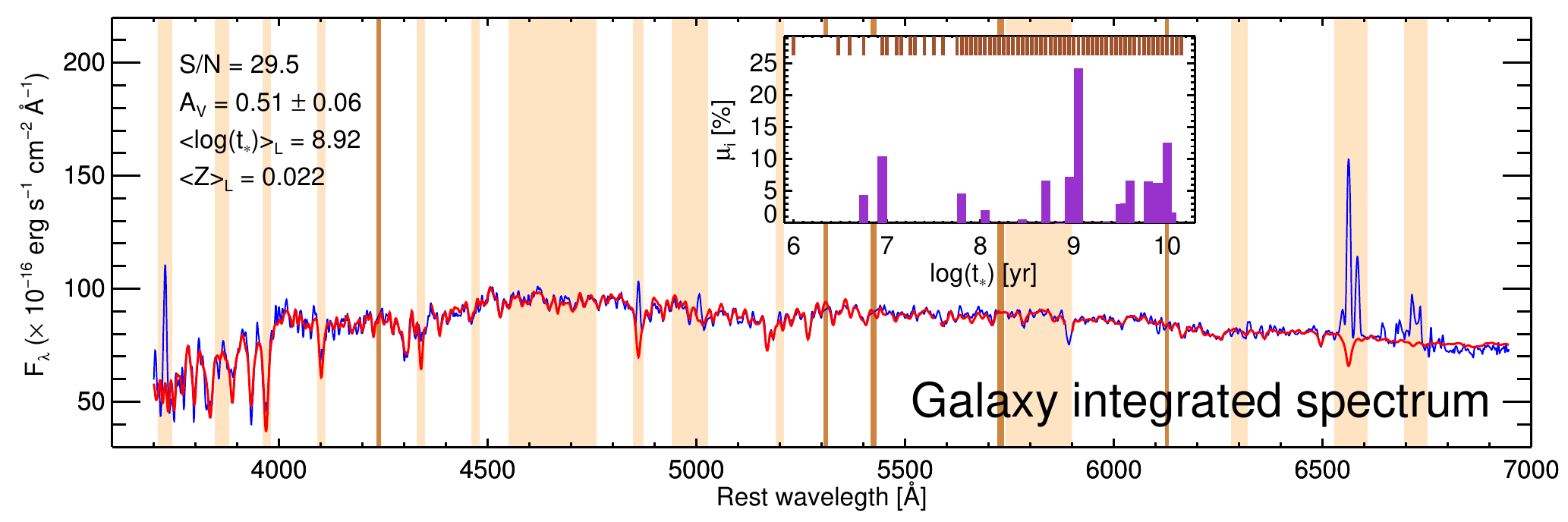}
\includegraphics[trim=-1cm 0cm -0.4cm 0cm,clip=true,width=\textwidth]{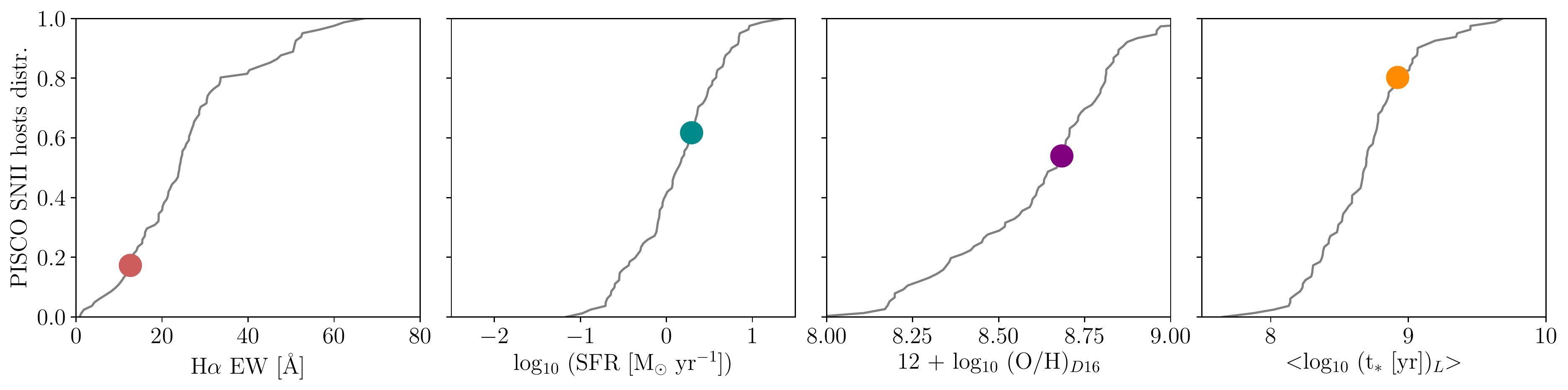}
\caption{{\it Top:} CCG 229-009 integrated spectrum and best simple stellar population fit from STARLIGHT (red). Vertical strips represent masked regions in the fit due to known emission lines (such as Balmer lines, oxygen, WR bumps, etc.; in beige) and regions with night-sky lines (in brown). The inner panel shows the star-formation history of the spectrum (in purple), where upper brown tick marks represent the ages of the models used in the fit. The signal-to-noise ratio of the spectrum, optical extinction, average stellar age, and average metallicity are reported in the upper-left corners.
{\it Bottom row:} Distributions of the same parameters of all 82 SN~II host galaxies in PISCO (Galbany et al. in prep.). Dotted colours represents the position of CGCG 229-009 in these distributions.}
\label{fig:ssp_global}
\end{figure*}

\begin{figure}
\includegraphics[width=\columnwidth]{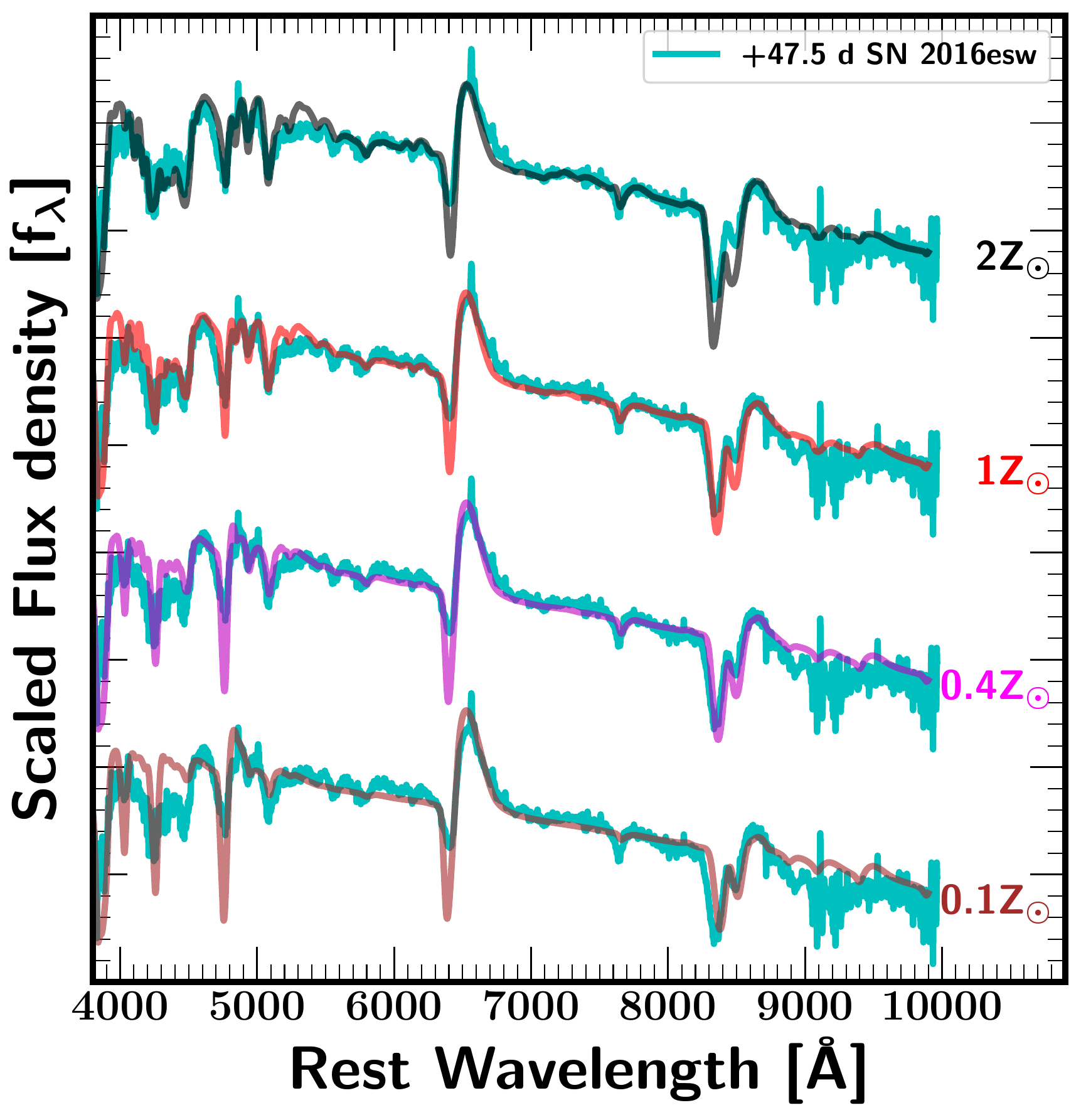}
\caption{Comparison theoretical spectra \citep{dessart14} and SN~2016esw observed spectrum. Four model spectra from progenitor models at 50 days post explosion with different metallicity are used: 0.1 (brown), 0.4 (magenta), 1 (red), and 2 (black) times solar metallicity. In cyan, the SN~2016esw spectrum at 47.5 days after the explosion corrected for host-galaxy and MWG extinctions is also shown.}
\label{fig:compa_Z}
\end{figure}

SN~II progenitors are massive stars ($M_\mathrm{{ZAMS}} \geq 8\,{\rm M}_{\odot}$; see \citealt{smartt09a}) that exploded a few tens of Myr after they were born predominantly within the dense cores of giant molecular clouds (surrounded by H~II regions).
From the 2D H$\alpha$ emission-line map, we detected 25 H~II regions and assigned the closest in distance to SN~2016esw as the ``parent'' H~II region, with the assumption that its progenitor was born there and shares the region's properties.
We present the spectrum of the SN parent H~II region in the top panel of Figure \ref{fig:ssp_local}, while in the bottom panel we show normalised cumulative distributions of 4 parameters: H$\alpha$ equivalent width (proxy for age; \citealt{kuncarayakti13,galbany14}); the star-formation rate (SFR), which is proportional to the H$\alpha$ luminosity \citep{kennicutt98}; the oxygen abundance as a proxy for the metal content (in the \citealt{dopita16} D16 calibrator); and the luminosity-weighted average stellar age. These four parameters are measured in the spectra of the 25 H~II regions found in CGCG 229-09. In each panel, we highlight where the SN parent H~II region is located in these distributions.

As seen in the top panel of Figure \ref{fig:ssp_local} where the star formation history (SFH) is displayed, STARLIGHT only needs three models of ages around 1~Gyr (600, 700, and 1100~Myr) to get the best fit to the spectrum.
However, the presence of emission lines suggests ongoing star formation happening in the region. In fact, we show that this H~II region has the highest H$\alpha$ EW of all H~II regions in the galaxy (73.82 $\pm$ 2.64 \AA).
This, combined with its low-to-average SFR (log$_{10}$[SFR/(M$_{\odot}$ yr$^{-1}$)] $= -1.56 \pm 0.06$~dex) compared to all H~II regions in the galaxy, can be explained by a projection effect where a small amount of gas (given the low/average SFR) is being ionised by young stars in the foreground, on top of an older stellar continuum in the background. 
This is reinforced by the fact that even having a low/average SFR, this H~II region has an older luminosity-weighted average age (log$_{10}(t/{\rm yr}) = 8.94 \pm 0.14$~dex) than their counterparts (see lower-right panel in Figure \ref{fig:ssp_local}).

The optical extinction in the local SN~2016esw environment can be estimated through the Balmer decrement \citep{stasinska04} in the continuum-subtracted spectrum. Assuming Case B recombination (typical of heating sources with $T \approx 10^4$~K and large optical depths; \citealt{osterbrock06}), we find $A_{\rm Vh} = 0.55 \pm 0.08$ mag (i.e., $E(B-V)\approx 0.18$ mag). Although this is totally independent of our previous estimate using the \ion{Na}{i}~D EW (see Section \ref{txt:avh}), all of the values are very consistent. This confirms the host-galaxy extinction adopted in this work, $E(B-V) \approx 0.185$ mag.

The luminosity-weighted stellar metallicity is above solar (Z$_\odot = 0.02$) with a value $<Z>_L = 0.03$ (i.e., 1.5\,Z$_{\odot}$), and it is in agreement with the estimated oxygen abundance, 12 $+$ log$_{10}$(O/H) = 8.71~dex. However, note that the luminosity-weighted stellar metallicity tells us the chemical composition when the stars were formed, not the current one of the gas. Since most of the population was formed at very early times (see Figure \ref{fig:ssp_local}), which is compatible with our current understanding of the SFH in galaxies, then the stellar metallicity should be lower than the current gas-phase one (unless there is a huge amount of accretion). With respect to the other H~II regions in the host galaxy, the SN parent H~II region has an average oxygen abundance value. 

A metallicity estimate can also be directly obtained using the SN spectra, as the strengths of metal lines are strongly dependent on progenitor metallicity \citep{dessart14}. In Figure \ref{fig:compa_Z}, we compare four theoretical models \citep{dessart14} with distinct progenitor metallicities (0.1, 0.4, 1, and 2 times Z$_\odot$) at 50 days after the explosion together with the SN~2016esw spectrum at 47.5 days after the explosion. We clearly see that the strengths of metal lines in the blue part of the SN~2016esw spectrum are a better fit to the high-metallicity models (1--2\,Z$_{\odot}$) than to those of low metallicity (0.1--0.4\,Z$_{\odot}$). This is consistent with the value derived from the SN parent H~II region, where a metallicity of 1.5\,Z$_{\odot}$ is found. 

One could suggest a possible relation between the peak luminosity and the metallicity; however, \citet{anderson16a} did not find such a correlation. On the other hand, \citet{taddia16a} found that SN~II peak magnitudes correlate with $Z$, but in the sense that brighter SNe~II may occur at lower $Z$, which is opposite what we found with our object (luminous SN~II and high-metallicity progenitor). Note also from this figure that the SN~2016esw spectrum corrected for MWG and host-galaxy extinctions fits perfectly the unextinguished theoretical models, again supporting the host-galaxy extinction value we derived.

Additionally, we present the integrated spectrum of CGCG 229-009 in Figure \ref{fig:ssp_global} and the best SSP fit from STARLIGHT. The lower panels show normalised cumulative distributions of the same four parameters presented in Figure \ref{fig:ssp_local} measured in integrated spectra of all 82 SN~II host galaxies from PISCO. The coloured dots represent the position of CGCG 229-009 within these distributions.

From the SFH shown in Figure \ref{fig:ssp_global} (top panel), we see that STARLIGHT needs a wide variety of SSPs of different ages (between 10~Myr and 10~Gyr), although the luminosity-weighted average age ($8.92 \pm 1.23$~dex) is quite similar to the SN~2016esw parent H~II region ($8.94 \pm 0.14$~dex).
From the lower panel of Figure \ref{fig:ssp_global}, we see that CGCG 229-009 is in the 20th percentile of old SN hosts in the PISCO sample, consistent with being in the 20th percentile of the galaxies with lower average H$\alpha$ EW ($12.63 \pm 1.40$ \AA) as both quantities correlate.

Finally, both the SFR ($0.29 \pm 0.43$~dex) and the oxygen abundance (8.68~dex) are very close to the average values for all PISCO SN~II hosts, although the SFR is a bit higher than the mean.
The main differences between our local and global environment analyses come from the H$\alpha$ EW, which is relatively high when looking at the SN position, but low when studying the global properties.

\section{Conclusions}

In this paper, for the first time, we show photometric and spectroscopic data on SN~2016esw, together with a study of its host galaxy. Thanks to our rapid follow-up observations from less than one day after the explosion up to 120 days later, and to a comparison with other SNe~II from the literature, we find a number of interesting similarities with the family of luminous SNe~II, as well as some peculiar properties that help us understand the great SN~II variety in the Universe. Our detailed observations should provide useful comparisons with models of SNe~II. We summarise the properties of SN~2016esw as follows.

\begin{enumerate}

\item{SN~2016esw is a luminous SNe~II, with an absolute magnitude ($M_V$) at maximum brightness between $-17.79$ and $-18.36$ mag (depending on the assumed host-galaxy extinction). As for other unusually luminous SNe~II (SN~2007pk or SN~2013by), there might be a contribution from CSM-ejecta interaction.\\}

\item{Ejecta-CSM interaction is also suggested by the early-time spectra ($<6.5$\,d) showing a nearly featureless blue continuum with few narrow emission lines, and by the boxy H$\alpha$ profile at epoch 19.5\,d. The mostly identical spectroscopic evolution with SN~2007pk, known to be characterised by strong ejecta-CSM interaction, also favours such an interpretation.\\}

\item{The early-time spectra (0.6\,d, 5.5\,d, and 6.5\,d) do not exhibit strong high-ionisation emission lines as seen in SNe~II having progenitors closely surrounded by dense CSM (e.g., SN~2013fs).\\}

\item{During the transition between an interacting to a normal SN~II (19.5 to 33.5 days), SN~2016esw spectra exhibit an H$\alpha$ P-Cygni profile with an atypical shape produced by an extra absorption component on the blue side associated with \ion{Si}{II}.\\}

\item{SN~2016esw also has an unusually long, flat, plateau slope relative to its luminosity; it seems to not follow the well-known relation between the luminosity and the plateau decline rate \citep{anderson14a}.\\}

\item{From the early-time light curves, the derived rise times are consistent with those from studies of large SN~II samples, and comparisons with analytical models lead to a small progenitor radius ($\sim 200$ ${\rm R}_{\odot}$).\\}

\item{Using host-galaxy information, a high-metallicity progenitor ($Z \approx 1.5$\,Z$_{\odot}$) is derived. SN~2016esw does not follow the correlation found by \citet{taddia16a} that more luminous SNe~II occur at lower $Z$.}

\end{enumerate}

\section*{Acknowledgements}

We thank the anonymous referee for useful suggestions that helped improve the quality of this paper.
Support for A.V.F.'s supernova research group at U.C. Berkeley has been provided by U.S. NSF grant AST-1211916, the TABASGO Foundation, Gary and Cynthia Bengier (T.d.J. is a Bengier Postdoctoral Fellow), the Christopher R. Redlich Fund, and the Miller Institute for Basic Research in Science (U.C. Berkeley). L.G. was supported in part by the NSF under grant AST-1311862. C.P.G. acknowledges support from EU/FP7-ERC grant No. 615929. The U.C.S.C. team is supported in part by NSF grant AST--1518052, the Gordon \& Betty Moore Foundation, and from fellowships from the Alfred P.\ Sloan Foundation and the David and Lucile Packard Foundation to R.J.F.

We are grateful to the following U.C. Berkeley undergraduate Nickel 1-m observers for their valuable help with this work: Nick Chocksi, Andrew Halle, Kevin Hayakawa, Andrew Rikhter, and Heechan Yuk. We thank the Lick Observatory staff for their expert assistance. KAIT and its ongoing operation were made possible by donations from Sun
Microsystems, Inc., the Hewlett-Packard Company, AutoScope Corporation, 
Lick Observatory, the NSF, the University of California, the Sylvia \& 
Jim Katzman Foundation, and the TABASGO Foundation.    
A major upgrade of the Kast spectrograph on the Shane 3~m telescope at 
Lick Observatory was made possible through generous gifts from the
Heising-Simons Foundation as well as William and Marina Kast.
Research at Lick Observatory is partially supported by a generous gift from Google. 
We also greatly appreciate contributions from numerous individuals, including Eliza Brown and Hal Candee,
Kathy Burck and Gilbert Montoya, David and Linda Cornfield, William and Phyllis Draper, Luke Ellis and Laura Sawczuk, Alan and Gladys Hoefer, Roger and Jody Lawler, DuBose and Nancy Montgomery, Rand Morimoto and Ana Henderson, Jeanne and Sanford Robertson,
Stanley and Miriam Schiffman, Thomas and Alison Schneider, Mary-Lou Smulders and Nicholas Hodson, Hans Spiller, Alan and Janet Stanford, the Hugh Stuart Center Charitable Trust, Clark and Sharon Winslow, Weldon and Ruth Wood, and many others. This research was made possible in part through the use of the AAVSO Photometric All-Sky Survey (APASS), funded by the Robert Martin Ayers Sciences Fund.





\bsp	
\label{lastpage}

\begin{thebibliography}{}
\makeatletter
\relax
\def\mn@urlcharsother{\let\do\@makeother \do\$\do\&\do\#\do\^\do\_\do\%\do\~}
\def\mn@doi{\begingroup\mn@urlcharsother \@ifnextchar [ {\mn@doi@}
  {\mn@doi@[]}}
\def\mn@doi@[#1]#2{\def\@tempa{#1}\ifx\@tempa\@empty \href
  {http://dx.doi.org/#2} {doi:#2}\else \href {http://dx.doi.org/#2} {#1}\fi
  \endgroup}
\def\mn@eprint#1#2{\mn@eprint@#1:#2::\@nil}
\def\mn@eprint@arXiv#1{\href {http://arxiv.org/abs/#1} {{\tt arXiv:#1}}}
\def\mn@eprint@dblp#1{\href {http://dblp.uni-trier.de/rec/bibtex/#1.xml}
  {dblp:#1}}
\def\mn@eprint@#1:#2:#3:#4\@nil{\def\@tempa {#1}\def\@tempb {#2}\def\@tempc
  {#3}\ifx \@tempc \@empty \let \@tempc \@tempb \let \@tempb \@tempa \fi \ifx
  \@tempb \@empty \def\@tempb {arXiv}\fi \@ifundefined
  {mn@eprint@\@tempb}{\@tempb:\@tempc}{\expandafter \expandafter \csname
  mn@eprint@\@tempb\endcsname \expandafter{\@tempc}}}

\bibitem[\protect\citeauthoryear{{Anderson}, {Gonz{\'a}lez-Gait{\'a}n}, {Hamuy}
   et~al.}{{Anderson} et~al.}{2014}]{anderson14a}
{Anderson} J.~P.,  {Gonz{\'a}lez-Gait{\'a}n} S.,  {Hamuy} M.,   et~al., 2014,
  \mn@doi [\apj] {10.1088/0004-637X/786/1/67}, \href
  {http://adsabs.harvard.edu/abs/2014ApJ...786...67A} {786, 67}

\bibitem[\protect\citeauthoryear{{Anderson}, {Gutierrez}, {Dessart}
  et~al.}{{Anderson} et~al.}{2016}]{anderson16a}
{Anderson} J.~P.,  {Gutierrez} C.~P.,  {Dessart} L.,   et~al., 2016, \mn@doi
  [\aap] {10.1051/0004-6361/201527691}, \href
  {http://adsabs.harvard.edu/abs/2016A%26A...589A.110A} {589, A110}

\bibitem[\protect\citeauthoryear{{Barbon}, {Ciatti}  \& {Rosino}}{{Barbon}
  et~al.}{1979}]{barbon79}
{Barbon} R.,  {Ciatti} F.,   {Rosino} L.,  1979, \aap, \href
  {http://adsabs.harvard.edu/abs/1979A%26A....72..287B} {72, 287}

\bibitem[\protect\citeauthoryear{{Baron} et~al.,}{{Baron}
  et~al.}{2000}]{baron00}
{Baron} E.,  et~al., 2000, \mn@doi [\apj] {10.1086/317795}, \href
  {http://adsabs.harvard.edu/abs/2000ApJ...545..444B} {545, 444}

\bibitem[\protect\citeauthoryear{{Bose}, {Sutaria}, {Kumar}  et~al.}{{Bose}
  et~al.}{2015}]{bose15}
{Bose} S.,  {Sutaria} F.,  {Kumar} B.,   et~al., 2015, \mn@doi [\apj]
  {10.1088/0004-637X/806/2/160}, \href
  {http://adsabs.harvard.edu/abs/2015ApJ...806..160B} {806, 160}

\bibitem[\protect\citeauthoryear{{Branch}, {Falk}, {Uomoto}, {Wills}, {McCall}
  \& {Rybski}}{{Branch} et~al.}{1981}]{bra81}
{Branch} D.,  {Falk} S.~W.,  {Uomoto} A.~K.,  {Wills} B.~J.,  {McCall} M.~L.,
  {Rybski} P.,  1981, \mn@doi [\apj] {10.1086/158755}, \href
  {http://adsabs.harvard.edu/abs/1981ApJ...244..780B} {244, 780}

\bibitem[\protect\citeauthoryear{{Cardelli}, {Clayton}  \& {Mathis}}{{Cardelli}
  et~al.}{1989}]{car89}
{Cardelli} J.~A.,  {Clayton} G.~C.,   {Mathis} J.~S.,  1989, \mn@doi [\apj]
  {10.1086/167900}, \href {http://adsabs.harvard.edu/abs/1989ApJ...345..245C}
  {345, 245}

\bibitem[\protect\citeauthoryear{{Chauvenet}}{{Chauvenet}}{1863}]{chauvenet1863}
{Chauvenet} W.,  1863, {A Manual of Spherical and Practical Astronomy V. II}

\bibitem[\protect\citeauthoryear{{Chevalier}}{{Chevalier}}{1976}]{chevalier76}
{Chevalier} R.~A.,  1976, \mn@doi [\apj] {10.1086/154557}, \href
  {http://adsabs.harvard.edu/abs/1976ApJ...207..872C} {207, 872}

\bibitem[\protect\citeauthoryear{{Chevalier}}{{Chevalier}}{1981}]{che81}
{Chevalier} R.~A.,  1981, \mn@doi [\apj] {10.1086/159460}, \href
  {http://adsabs.harvard.edu/abs/1981ApJ...251..259C} {251, 259}

\bibitem[\protect\citeauthoryear{{Cid Fernandes}, {Mateus}, {Sodr{\'e}},
  {Stasi{\'n}ska}  \& {Gomes}}{{Cid Fernandes} et~al.}{2005}]{cidfernandes05}
{Cid Fernandes} R.,  {Mateus} A.,  {Sodr{\'e}} L.,  {Stasi{\'n}ska} G.,
  {Gomes} J.~M.,  2005, \mn@doi [\mnras] {10.1111/j.1365-2966.2005.08752.x},
  \href {http://adsabs.harvard.edu/abs/2005MNRAS.358..363C} {358, 363}


\bibitem[\protect\citeauthoryear{{Colgate}}{{Colgate}}{1974}]{colgate74}
{Colgate} S.~A.,  1974, \mn@doi [\apj] {10.1086/152632}, \href
  {http://adsabs.harvard.edu/abs/1974ApJ...187..333C} {187, 333}

\bibitem[\protect\citeauthoryear{{Cowen}, {Franckowiak}  \& {Kowalski}}{{Cowen}
  et~al.}{2010}]{cowen10}
{Cowen} D.~F.,  {Franckowiak} A.,   {Kowalski} M.,  2010, \mn@doi
  [Astroparticle Physics] {10.1016/j.astropartphys.2009.10.007}, \href
  {http://adsabs.harvard.edu/abs/2010APh....33...19C} {33, 19}

\bibitem[\protect\citeauthoryear{{D'Andrea}, {Sako}, {Dilday}
  et~al.}{{D'Andrea} et~al.}{2010}]{andrea10}
{D'Andrea} C.~B.,  {Sako} M.,  {Dilday} B.,   et~al., 2010, \mn@doi [\apj]
  {10.1088/0004-637X/708/1/661}, \href
  {http://adsabs.harvard.edu/abs/2010ApJ...708..661D} {708, 661}

\bibitem[\protect\citeauthoryear{{de Jaeger}, {Gonz{\'a}lez-Gait{\'a}n},
  {Anderson}  et~al.}{{de Jaeger} et~al.}{2015}]{dejaeger15b}
{de Jaeger} T.,  {Gonz{\'a}lez-Gait{\'a}n} S.,  {Anderson} J.~P.,   et~al.,
  2015, \mn@doi [\apj] {10.1088/0004-637X/815/2/121}, \href
  {http://adsabs.harvard.edu/abs/2015ApJ...815..121D} {815, 121}

\bibitem[\protect\citeauthoryear{{de Jaeger}, {Galbany}, {Filippenko}
  et~al.}{{de Jaeger} et~al.}{2017a}]{dejaeger17b}
{de Jaeger} T.,  {Galbany} L.,  {Filippenko} A.~V.,   et~al., 2017a, \mn@doi
  [\mnras] {10.1093/mnras/stx2300}, \href
  {http://adsabs.harvard.edu/abs/2017MNRAS.472.4233D} {472, 4233}

\bibitem[\protect\citeauthoryear{{de Jaeger}, {Gonz{\'a}lez-Gait{\'a}n},
  {Hamuy}  et~al.}{{de Jaeger} et~al.}{2017b}]{dejaeger17a}
{de Jaeger} T.,  {Gonz{\'a}lez-Gait{\'a}n} S.,  {Hamuy} M.,   et~al., 2017b,
  \mn@doi [\apj] {10.3847/1538-4357/835/2/166}, \href
  {http://adsabs.harvard.edu/abs/2017ApJ...835..166D} {835, 166}

\bibitem[\protect\citeauthoryear{{de Jaeger}, {Anderson},{Galbany},
  et~al.}{{de Jaeger} et~al.}{2018a}]{dejaeger18a}
{de Jaeger} T.,  {Anderson} J.~P., {Galbany} L.,   et~al., 2018, \mn@doi
  [\mnras] {10.1093/mnras/sty508}, \href
  {http://adsabs.harvard.edu/abs/2018MNRAS.tmp..487D} {487}

\bibitem[\protect\citeauthoryear{{Dessart}, {Gutierrez}, {Hamuy}
  et~al.}{{Dessart} et~al.}{2014}]{dessart14}
{Dessart} L.,  {Gutierrez} C.~P.,  {Hamuy} M.,   et~al., 2014, \mn@doi [\mnras]
  {10.1093/mnras/stu417}, \href
  {http://adsabs.harvard.edu/abs/2014MNRAS.440.1856D} {440, 1856}

\bibitem[\protect\citeauthoryear{{Dessart}, {Hillier}, and {Audit}}{{Dessart} et~al.}{2017}]{dessart17}
{Dessart}, L. and {Hillier}, D.~J. and {Audit}, E.,   et~al., 2017, \mn@doi [\aap]
  {10.1051/0004-6361/201730942}, \href
  {http://adsabs.harvard.edu/abs/2017A%26A...605A..83D} {605, 83}

\bibitem[\protect\citeauthoryear{{de Vaucouleurs}, {de Vaucouleurs}, {Buta},
  {Ables}  \& {Hewitt}}{{de Vaucouleurs} et~al.}{1981}]{devaucouleurs81}
{de Vaucouleurs} G.,  {de Vaucouleurs} A.,  {Buta} R.,  {Ables} H.~D.,
  {Hewitt} A.~V.,  1981, \mn@doi [\pasp] {10.1086/130772}, \href
  {http://adsabs.harvard.edu/abs/1981PASP...93...36D} {93, 36}


\bibitem[\protect\citeauthoryear{{Dhungana}, {Kehoe}, {Vinko}
  et~al.}{{Dhungana} et~al.}{2016}]{dhungana16}
{Dhungana} G.,  {Kehoe} R.,  {Vinko} J.,   et~al., 2016, \mn@doi [\apj]
  {10.3847/0004-637X/822/1/6}, \href
  {http://adsabs.harvard.edu/abs/2016ApJ...822....6D} {822, 6}

\bibitem[\protect\citeauthoryear{{Dopita}, {Kewley}, {Sutherland}  \&
  {Nicholls}}{{Dopita} et~al.}{2016}]{dopita16}
{Dopita} M.~A.,  {Kewley} L.~J.,  {Sutherland} R.~S.,   {Nicholls} D.~C.,
  2016, \mn@doi [\apss] {10.1007/s10509-016-2657-8}, \href
  {http://adsabs.harvard.edu/abs/2016Ap%26SS.361...61D} {361, 61}

\bibitem[\protect\citeauthoryear{{Elias-Rosa}, {Van Dyk}, {Li}
  et~al.}{{Elias-Rosa} et~al.}{2010}]{EliasRosa10}
{Elias-Rosa} N.,  {Van Dyk} S.~D.,  {Li} W.,   et~al., 2010, \mn@doi [\apjl]
  {10.1088/2041-8205/714/2/L254}, \href
  {http://adsabs.harvard.edu/abs/2010ApJ...714L.254E} {714, L254}

\bibitem[\protect\citeauthoryear{{Elias-Rosa}, {Van Dyk}, {Li}
  et~al.}{{Elias-Rosa} et~al.}{2011}]{EliasRosa11}
{Elias-Rosa} N.,  {Van Dyk} S.~D.,  {Li} W.,   et~al., 2011, \mn@doi [\apj]
  {10.1088/0004-637X/742/1/6}, \href
  {http://adsabs.harvard.edu/abs/2011ApJ...742....6E} {742, 6}

\bibitem[\protect\citeauthoryear{{Elmhamdi}, {Danziger}, {Chugai}
  et~al.}{{Elmhamdi} et~al.}{2003}]{elmhamdi03}
{Elmhamdi} A.,  {Danziger} I.~J.,  {Chugai} N.,   et~al., 2003, \mn@doi
  [\mnras] {10.1046/j.1365-8711.2003.06150.x}, \href
  {http://adsabs.harvard.edu/abs/2003MNRAS.338..939E} {338, 939}

\bibitem[\protect\citeauthoryear{{Falk} \& {Arnett}}{{Falk} \&
  {Arnett}}{1977}]{falk77}
{Falk} S.~W.,  {Arnett} W.~D.,  1977, \mn@doi [\apjs] {10.1086/190440}, \href
  {http://adsabs.harvard.edu/abs/1977ApJS...33..515F} {33, 515}

\bibitem[\protect\citeauthoryear{{Faran}, {Poznanski}, {Filippenko}
  et~al.}{{Faran} et~al.}{2014a}]{faran14a}
{Faran} T.,  {Poznanski} D.,  {Filippenko} A.~V.,   et~al., 2014a, \mn@doi
  [\mnras] {10.1093/mnras/stu955}, \href
  {http://adsabs.harvard.edu/abs/2014MNRAS.442..844F} {442, 844}

\bibitem[\protect\citeauthoryear{{Faran}, {Poznanski}, {Filippenko}
  et~al.}{{Faran} et~al.}{2014b}]{faran14b}
{Faran} T.,  {Poznanski} D.,  {Filippenko} A.~V.,   et~al., 2014b, \mn@doi
  [\mnras] {10.1093/mnras/stu1760}, \href
  {http://adsabs.harvard.edu/abs/2014MNRAS.445..554F} {445, 554}

\bibitem[\protect\citeauthoryear{{Faran}, {Nakar}  \& {Poznanski}}{{Faran}
  et~al.}{2017}]{faran17}
{Faran} T.,  {Nakar} E.,   {Poznanski} D.,  2017, preprint, \href
  {http://adsabs.harvard.edu/abs/2017arXiv170707695F} {} (\mn@eprint {arXiv}
  {1707.07695})

\bibitem[\protect\citeauthoryear{{Filippenko}}{{Filippenko}}{1982}]{filippenko82}
{Filippenko} A.~V.,  1982, \mn@doi [\pasp] {10.1086/131052}, \href
  {http://adsabs.harvard.edu/abs/1982PASP...94..715F} {94, 715}

\bibitem[\protect\citeauthoryear{{Filippenko}}{{Filippenko}}{1988}]{filippenko88}
{Filippenko} A.~V.,  1988, \mn@doi [\aj] {10.1086/114940}, \href
  {http://adsabs.harvard.edu/abs/1988AJ.....96.1941F} {96, 1941}

\bibitem[\protect\citeauthoryear{{Filippenko}}{{Filippenko}}{1991}]{filippenko91}
{Filippenko} A.~V.,  1991, in {Danziger} I.~J.,  {Kjaer} K.,  eds,  European
  Southern Observatory Conference and Workshop Proceedings Vol. 37, European
  Southern Observatory Conference and Workshop Proceedings. p.~343

\bibitem[\protect\citeauthoryear{{Filippenko}}{{Filippenko}}{1997}]{filippenko97}
{Filippenko} A.~V.,  1997, \mn@doi [\araa] {10.1146/annurev.astro.35.1.309},
  \href {http://adsabs.harvard.edu/abs/1997ARA%26A..35..309F} {35, 309}

\bibitem[\protect\citeauthoryear{{Filippenko}, {Matheson}  \&
  {Ho}}{{Filippenko} et~al.}{1993}]{filippenko93}
{Filippenko} A.~V.,  {Matheson} T.,   {Ho} L.~C.,  1993, \mn@doi [\apjl]
  {10.1086/187043}, \href {http://adsabs.harvard.edu/abs/1993ApJ...415L.103F}
  {415, L103}

\bibitem[\protect\citeauthoryear{{Filippenko}, {Li}, {Treffers}  \&
  {Modjaz}}{{Filippenko} et~al.}{2001}]{filippenko01}
{Filippenko} A.~V.,  {Li} W.~D.,  {Treffers} R.~R.,   {Modjaz} M.,  2001, in
  {Paczynski} B.,  {Chen} W.-P.,   {Lemme} C.,  eds,  Astronomical Society of
  the Pacific Conference Series Vol. 246, IAU Colloq. 183: Small Telescope
  Astronomy on Global Scales. p.~121

\bibitem[\protect\citeauthoryear{{Fransson}}{{Fransson}}{1982}]{fra82}
{Fransson} C.,  1982, \aap, \href
  {http://adsabs.harvard.edu/abs/1982A%26A...111..140F} {111, 140}

\bibitem[\protect\citeauthoryear{{Freedman}, {Madore}, {Gibson}
  et~al.}{{Freedman} et~al.}{2001}]{freedman01}
{Freedman} W.~L.,  {Madore} B.~F.,  {Gibson} B.~K.,   et~al., 2001, \mn@doi
  [\apj] {10.1086/320638}, \href
  {http://adsabs.harvard.edu/abs/2001ApJ...553...47F} {553, 47}

\bibitem[\protect\citeauthoryear{{Gal-Yam}, {Arcavi}, {Ofek}  et~al.}{{Gal-Yam}
  et~al.}{2014}]{galyam14}
{Gal-Yam} A.,  {Arcavi} I.,  {Ofek} E.~O,   et~al., 2014, \mn@doi [Nature] {10.1038/nature13304}, \href
  {http://adsabs.harvard.edu/abs/2014Natur.509..471G} {509, 471}


\bibitem[\protect\citeauthoryear{{Galbany}, {Stanishev},{Mour{\~a}o}}{{Galbany}
  et~al.}{2014}]{galbany14}
{Galbany} L., {Stanishev}, V., {Mour{\~a}o}, A.~M., et~al., 2014, \mn@doi [\aap] {10.1051/0004-6361/201424717},
  \href {http://adsabs.harvard.edu/abs/2014A%26A...572A..38G} {572, A38}

\bibitem[\protect\citeauthoryear{{Galbany}, {Hamuy}, {Phillips}
  et~al.}{{Galbany} et~al.}{2016a}]{galbany16a}
{Galbany} L.,  {Hamuy} M.,  {Phillips} M.~M.,   et~al., 2016a, \mn@doi [\aj]
  {10.3847/0004-6256/151/2/33}, \href
  {http://adsabs.harvard.edu/abs/2016AJ....151...33G} {151, 33}

\bibitem[\protect\citeauthoryear{{Galbany}, {Stanishev}, {Mour{\~a}o}}{{Galbany}
  et~al.}{2016b}]{galbany16b}
{Galbany} L., {Stanishev}, V., {Mour{\~a}o}, A.~M.  et~al., 2016b, \mn@doi [\aap] {10.1051/0004-6361/201528045},
  \href {http://adsabs.harvard.edu/abs/2016A%26A...591A..48G} {591, A48}

\bibitem[\protect\citeauthoryear{{Galbany}, {Mora}, {Gonz{\'a}lez-Gait{\'a}n}
  et~al.}{{Galbany} et~al.}{2017}]{galbany17}
{Galbany} L.,  {Mora} L.,  {Gonz{\'a}lez-Gait{\'a}n} S.,   et~al., 2017,
  \mn@doi [\mnras] {10.1093/mnras/stx367}, \href
  {http://adsabs.harvard.edu/abs/2017MNRAS.468..628G} {468, 628}

\bibitem[\protect\citeauthoryear{{Galbany}, {Anderson}, {S\'anchez}
  et~al.}{{Galbany} et~al.}{2018}]{galbany18}
{Galbany} L.,  {Anderson} J. P.,  {S\'anchez} S. F.,   et~al., 2018,
  \mn@doi [\apj] {10.3847/1538-4357/aaaf20}, \href
  {http://adsabs.harvard.edu/abs/2018ApJ...855..107G} {855, 107}

\bibitem[\protect\citeauthoryear{{Gall}, {Polshaw}, {Kotak}  et~al.}{{Gall}
  et~al.}{2015}]{gall15}
{Gall} E.~E.~E.,  {Polshaw} J.,  {Kotak} R.,   et~al., 2015, \mn@doi [\aap]
  {10.1051/0004-6361/201525868}, \href
  {http://adsabs.harvard.edu/abs/2015A%26A...582A...3G} {582, A3}

\bibitem[\protect\citeauthoryear{{Ganeshalingam}, {Li}, {Filippenko}
  et~al.}{{Ganeshalingam} et~al.}{2010}]{ganeshalingam10}
{Ganeshalingam} M.,  {Li} W.,  {Filippenko} A.~V.,   et~al., 2010, \mn@doi
  [\apjs] {10.1088/0067-0049/190/2/418}, \href
  {http://adsabs.harvard.edu/abs/2010ApJS..190..418G} {190, 418}

\bibitem[\protect\citeauthoryear{{Gonz{\'a}lez-Gait{\'a}n}, {Tominaga},
  {Molina}  et~al.}{{Gonz{\'a}lez-Gait{\'a}n} et~al.}{2015}]{gonzalezgaitan14}
{Gonz{\'a}lez-Gait{\'a}n} S.,  {Tominaga} N.,  {Molina} J.,   et~al., 2015,
  \mn@doi [\mnras] {10.1093/mnras/stv1097}, \href
  {http://adsabs.harvard.edu/abs/2015MNRAS.451.2212G} {451, 2212}

\bibitem[\protect\citeauthoryear{{Grassberg}, {Imshennik}  \&
  {Nadyozhin}}{{Grassberg} et~al.}{1971}]{grassberg71}
{Grassberg} E.~K.,  {Imshennik} V.~S.,   {Nadyozhin} D.~K.,  1971, \mn@doi
  [\apss] {10.1007/BF00654604}, \href
  {http://adsabs.harvard.edu/abs/1971Ap%26SS..10...28G} {10, 28}

\bibitem[\protect\citeauthoryear{{Guillochon}, {Parrent}, {Kelley}  \&
  {Margutti}}{{Guillochon} et~al.}{2016}]{guillochon16}
{Guillochon} J.,  {Parrent} J.,  {Kelley} L.~Z.,   {Margutti} R.,  2016, \mn@doi
  [\apj] {10.3847/1538-4357/835/1/64}, \href
  {http://adsabs.harvard.edu/abs/2017ApJ...835...64G} {835, 64}

\bibitem[\protect\citeauthoryear{{Guti{\'e}rrez}, {Anderson}, {Hamuy}
  et~al.}{{Guti{\'e}rrez} et~al.}{2014}]{gutierrez14}
{Guti{\'e}rrez} C.~P.,  {Anderson} J.~P.,  {Hamuy} M.,   et~al., 2014, \mn@doi
  [\apjl] {10.1088/2041-8205/786/2/L15}, \href
  {http://adsabs.harvard.edu/abs/2014ApJ...786L..15G} {786, L15}

\bibitem[\protect\citeauthoryear{{Guti{\'e}rrez}, {Anderson}, {Hamuy}
  et~al.}{{Guti{\'e}rrez} et~al.}{2017}]{gutierrez17a}
{Guti{\'e}rrez} C.~P.,  {Anderson} J.~P.,  {Hamuy} M.,   et~al., 2017,
 \mn@doi [\apj] {10.3847/1538-4357/aa8f42}, \href
  {http://adsabs.harvard.edu/abs/2017ApJ...850...90G} {850, 90}


\bibitem[\protect\citeauthoryear{{Halevi}, {Zheng}  \& {Filippenko}}{{Halevi}
  et~al.}{2016}]{haveli16}
{Halevi} G.,  {Zheng} W.,   {Filippenko} A.~V.,  2016, Transient Name Server
  Discovery Report, \href {http://adsabs.harvard.edu/abs/2016TNSTR.542....1H}
  {542}

\bibitem[\protect\citeauthoryear{{Hamuy}}{{Hamuy}}{2001}]{hamuyphd}
{Hamuy} M.~A.,  2001, PhD thesis, The University of Arizona

\bibitem[\protect\citeauthoryear{{Hamuy}}{{Hamuy}}{2003}]{hamuy03a}
{Hamuy} M.,  2003, \mn@doi [\apj] {10.1086/344689}, \href
  {http://adsabs.harvard.edu/abs/2003ApJ...582..905H} {582, 905}

\bibitem[\protect\citeauthoryear{{Hamuy} \& {Pinto}}{{Hamuy} \&
  {Pinto}}{2002}]{hamuy02}
{Hamuy} M.,  {Pinto} P.~A.,  2002, \mn@doi [\apjl] {10.1086/339676}, \href
  {http://adsabs.harvard.edu/abs/2002ApJ...566L..63H} {566, L63}

\bibitem[\protect\citeauthoryear{{Hamuy}, {Pinto}, {Maza}  et~al.}{{Hamuy}
  et~al.}{2001}]{hamuy01}
{Hamuy} M.,  {Pinto} P.~A.,  {Maza} J.,   et~al., 2001, \mn@doi [\apj]
  {10.1086/322450}, \href {http://adsabs.harvard.edu/abs/2001ApJ...558..615H}
  {558, 615}

\bibitem[\protect\citeauthoryear{{Hicken} et~al.,}{{Hicken}
  et~al.}{2017}]{hicken17}
{Hicken} M.,  et~al., 2017, preprint, \href
  {http://adsabs.harvard.edu/abs/2017arXiv170601030H} {} (\mn@eprint {arXiv}
  {1706.01030})

\bibitem[\protect\citeauthoryear{{Huang}, {Wang}, {Zhang}  et~al.}{{Huang}
  et~al.}{2015}]{huang15}
{Huang} F.,  {Wang} X.,  {Zhang} J.,   et~al., 2015, \mn@doi [\apj]
  {10.1088/0004-637X/807/1/59}, \href
  {http://adsabs.harvard.edu/abs/2015ApJ...807...59H} {807, 59}

\bibitem[\protect\citeauthoryear{{Inserra}, {Turatto}, {Pastorello}
  et~al.}{{Inserra} et~al.}{2011}]{inserra11}
{Inserra} C.,  {Turatto} M.,  {Pastorello} A.,   et~al., 2011, \mn@doi [\mnras]
  {10.1111/j.1365-2966.2011.19128.x}, \href
  {http://adsabs.harvard.edu/abs/2011MNRAS.417..261I} {417, 261}

\bibitem[\protect\citeauthoryear{{Inserra}, {Turatto}, {Pastorello}
  et~al.}{{Inserra} et~al.}{2012}]{inserra12}
{Inserra} C.,  {Turatto} M.,  {Pastorello} A.,   et~al., 2012, \mn@doi [\mnras]
  {10.1111/j.1365-2966.2012.20685.x}, \href
  {http://adsabs.harvard.edu/abs/2012MNRAS.422.1122I} {422, 1122}

\bibitem[\protect\citeauthoryear{{Inserra} et~al.,}{{Inserra}
  et~al.}{2013}]{inserra13}
{Inserra} C.,  et~al., 2013, \mn@doi [\aap] {10.1051/0004-6361/201220496},
  \href {http://adsabs.harvard.edu/abs/2013A%26A...555A.142I} {555, A142}

\bibitem[\protect\citeauthoryear{{Kennicutt}}{{Kennicutt}}{1998}]{kennicutt98}
{Kennicutt} Jr. R.~C.,  1998, \mn@doi [\araa] {10.1146/annurev.astro.36.1.189},
  \href {http://adsabs.harvard.edu/abs/1998ARA%26A..36..189K} {36, 189}

\bibitem[\protect\citeauthoryear{{Khazov}, {Yaron}, {Gal-Yam}
  et~al.}{{Khazov} et~al.}{2016}]{khazov16}
{Khazov} D.,  {Yaron} O.,  {Gal-Yam} A.,   et~al., 2016,
  \mn@doi [\apj] {10.3847/0004-637X/818/1/3}, \href
  {http://adsabs.harvard.edu/abs/2016ApJ...818....3K} {818, 3}

\bibitem[\protect\citeauthoryear{{Klein} \& {Chevalier}}{{Klein} \&
  {Chevalier}}{1978}]{klein78}
{Klein} R.~I.,  {Chevalier} R.~A.,  1978, \mn@doi [\apjl] {10.1086/182740},
  \href {http://adsabs.harvard.edu/abs/1978ApJ...223L.109K} {223, L109}

\bibitem[\protect\citeauthoryear{{Kuncarayakti} et~al.,}{{Kuncarayakti}
  et~al.}{2013}]{kuncarayakti13}
{Kuncarayakti} H.,  et~al., 2013, \mn@doi [\aj] {10.1088/0004-6256/146/2/31},
  \href {http://adsabs.harvard.edu/abs/2013AJ....146...31K} {146, 31}

\bibitem[\protect\citeauthoryear{{Landolt}}{{Landolt}}{1992}]{landolt92}
{Landolt} A.~U.,  1992, \mn@doi [\aj] {10.1086/116242}, \href
  {http://adsabs.harvard.edu/abs/1992AJ....104..340L} {104, 340}

\bibitem[\protect\citeauthoryear{{Leaman}, {Li}, {Chornock}  \&
  {Filippenko}}{{Leaman} et~al.}{2011}]{leaman11}
{Leaman} J.,  {Li} W.,  {Chornock} R.,   {Filippenko} A.~V.,  2011, \mn@doi
  [\mnras] {10.1111/j.1365-2966.2011.18158.x}, \href
  {http://adsabs.harvard.edu/abs/2011MNRAS.412.1419L} {412, 1419}

\bibitem[\protect\citeauthoryear{{Leonard}, {Filippenko}, {Gates}
  et~al.}{{Leonard} et~al.}{2002}]{leonard02}
{Leonard} D.~C.,  {Filippenko} A.~V.,  {Gates} E.~L.,   et~al., 2002, \mn@doi
  [\pasp] {10.1086/324785}, \href
  {http://adsabs.harvard.edu/abs/2002PASP..114...35L} {114, 35}

\bibitem[\protect\citeauthoryear{{Li}, {Filippenko}, {Chornock}  \& {Jha}}{{Li}
  et~al.}{2003}]{li03}
{Li} W.,  {Filippenko} A.~V.,  {Chornock} R.,   {Jha} S.,  2003, \mn@doi
  [\pasp] {10.1086/376432}, \href
  {http://adsabs.harvard.edu/abs/2003PASP..115..844L} {115, 844}

\bibitem[\protect\citeauthoryear{{Maguire}, {Di Carlo}, {Smartt}, {Pastorello}
  et~al.}{{Maguire} et~al.}{2010}]{maguire10b}
{Maguire} K.,  {Di Carlo} E.,  {Smartt} S.~J.,  {Pastorello} A.,   et~al.,
  2010, \mn@doi [\mnras] {10.1111/j.1365-2966.2010.16332.x}, \href
  {http://adsabs.harvard.edu/abs/2010MNRAS.404..981M} {404, 981}

\bibitem[\protect\citeauthoryear{{Martini} et~al.,}{{Martini}
  et~al.}{2014}]{martini14}
{Martini} P.,  et~al., 2014, in Ground-based and Airborne Instrumentation for
  Astronomy V. p. 91470Z (\mn@eprint {arXiv} {1407.4541})


\bibitem[\protect\citeauthoryear{{Mast} et~al.,}{{Mast}
  et~al.}{2012}]{mast14}
{Mast, D., Rosales-Ortega, F.~F., S{\'a}nchez} S.~F.,  et~al., 2014, \mn@doi [\aap]
  {10.1051/0004-6361/201321789}, \href
  {http://adsabs.harvard.edu/abs/2014A%26A...561A.129M} {561, 129}


\bibitem[\protect\citeauthoryear{{Mauerhan}, {Van Dyk}, {Johansson}
  et~al.}{{Mauerhan} et~al.}{2016}]{mauerhan16}
{Mauerhan} J.~C.,  {Van Dyk} S.~D.,  {Johansson} J.,   et~al., 2016, \mn@doi [\apj]
{10.3847/1538-4357/834/2/118}, \href{http://adsabs.harvard.edu/abs/2017ApJ...834..118M}{834, 118}

\bibitem[\protect\citeauthoryear{{Miller} \& {Stone}}{{Miller} \&
  {Stone}}{1993}]{miller93}
{Miller} R.~P.~S.,  {Stone} R.~P.,  1993, Lick Observatory Technical Report 66
  (UC Santa Cruz)

\bibitem[\protect\citeauthoryear{{Morozova}, {Piro}  \& {Valenti}}{{Morozova}
  et~al.}{2017a}]{morozova17b}
{Morozova} V., {Piro} A.~L., {Valenti} S., 2017a, preprint, \href
  {http://adsabs.harvard.edu/abs/2017arXiv170904928M} {} (\mn@eprint {arXiv}
  {1709.04928})

\bibitem[\protect\citeauthoryear{{Morozova}, {Piro}  \& {Valenti}}{{Morozova}
  et~al.}{2017b}]{morozova17}
{Morozova} V., {Piro} A.~L., {Valenti} S., 2017b, \mn@doi [\apj]
  {10.3847/1538-4357/aa6251}, \href
  {http://adsabs.harvard.edu/abs/2017ApJ...838...28M} {838, 28}

\bibitem[\protect\citeauthoryear{{Nakar} \& {Sari}}{{Nakar} \&
  {Sari}}{2010}]{nakar10}
{Nakar} E., {Sari} R., 2010, \mn@doi [\apj] {10.1088/0004-637X/725/1/904},
  \href {http://adsabs.harvard.edu/abs/2010ApJ...725..904N} {725, 904}

\bibitem[\protect\citeauthoryear{{Nugent}, {Sullivan}, {Ellis}
  et~al.}{{Nugent} et~al.}{2006}]{nugent06}
{Nugent} P.,  {Sullivan} M.,  {Ellis} R.,   et~al., 2006, \mn@doi [\apj]
  {10.1086/504413}, \href {http://adsabs.harvard.edu/abs/2006ApJ...645..841N}
  {645, 841}

\bibitem[\protect\citeauthoryear{{Olivares E.}, {Hamuy}, {Pignata}
  et~al.}{{Olivares E.} et~al.}{2010}]{olivares10}
{Olivares E.} F.,  {Hamuy} M.,  {Pignata} G.,   et~al., 2010, \mn@doi [\apj]
  {10.1088/0004-637X/715/2/833}, \href
  {http://adsabs.harvard.edu/abs/2010ApJ...715..833O} {715, 833}

\bibitem[\protect\citeauthoryear{{Osterbrock} \& {Ferland}}{{Osterbrock} \&
  {Ferland}}{2006}]{osterbrock06}
{Osterbrock} D.~E.,  {Ferland} G.~J.,  2006, {Astrophysics of gaseous nebulae
  and active galactic nuclei}

\bibitem[\protect\citeauthoryear{{Parisky} \& {Li}}{{Parisky} \&
  {Li}}{2007}]{parisky07}
{Parisky} X.,  {Li} W.,  2007, Central Bureau Electronic Telegrams, \href
  {http://adsabs.harvard.edu/abs/2007CBET.1129....1P} {1129}

\bibitem[\protect\citeauthoryear{{Pastorello}, {Sauer}, {Taubenberger},
  {Mazzali}  et~al.}{{Pastorello} et~al.}{2006}]{pastorello06}
{Pastorello} A.,  {Sauer} D.,  {Taubenberger} S.,  {Mazzali} P.~A.,   et~al.,
  2006, \mn@doi [\mnras] {10.1111/j.1365-2966.2006.10587.x}, \href
  {http://adsabs.harvard.edu/abs/2006MNRAS.370.1752P} {370, 1752}

\bibitem[\protect\citeauthoryear{{Pastorello}, {Valenti}, {Zampieri}
  et~al.}{{Pastorello} et~al.}{2009}]{pastorello09}
{Pastorello} A.,  {Valenti} S.,  {Zampieri} L.,   et~al., 2009, \mn@doi
  [\mnras] {10.1111/j.1365-2966.2009.14505.x}, \href
  {http://adsabs.harvard.edu/abs/2009MNRAS.394.2266P} {394, 2266}

\bibitem[\protect\citeauthoryear{{Phillips}, {Simon}, {Morrell}
  et~al.}{{Phillips} et~al.}{2013}]{phillips13}
{Phillips} M.~M.,  {Simon} J.~D.,  {Morrell} N.,   et~al., 2013, \mn@doi [\apj]
  {10.1088/0004-637X/779/1/38}, \href
  {http://adsabs.harvard.edu/abs/2013ApJ...779...38P} {779, 38}

\bibitem[\protect\citeauthoryear{{Popov}}{{Popov}}{1993}]{popov93}
{Popov} D.~V.,  1993, \mn@doi [\apj] {10.1086/173117}, \href
  {http://adsabs.harvard.edu/abs/1993ApJ...414..712P} {414, 712}

\bibitem[\protect\citeauthoryear{{Poznanski}, {Nugent}  \&
  {Filippenko}}{{Poznanski} et~al.}{2010}]{poznanski10}
{Poznanski} D.,  {Nugent} P.~E.,   {Filippenko} A.~V.,  2010, \mn@doi [\apj]
  {10.1088/0004-637X/721/2/956}, \href
  {http://adsabs.harvard.edu/abs/2010ApJ...721..956P} {721, 956}

\bibitem[\protect\citeauthoryear{{Poznanski}, {Ganeshalingam}, {Silverman}  \&
  {Filippenko}}{{Poznanski} et~al.}{2011}]{poznanski11}
{Poznanski} D.,  {Ganeshalingam} M.,  {Silverman} J.~M.,   {Filippenko} A.~V.,
  2011, \mn@doi [\mnras] {10.1111/j.1745-3933.2011.01084.x}, \href
  {http://adsabs.harvard.edu/abs/2011MNRAS.415L..81P} {415, L81}

\bibitem[\protect\citeauthoryear{{Rabinak} \& {Waxman}}{{Rabinak} \&
  {Waxman}}{2011}]{rabinak11}
{Rabinak} I.,  {Waxman} E.,  2011, \mn@doi [\apj] {10.1088/0004-637X/728/1/63},
  \href {http://adsabs.harvard.edu/abs/2011ApJ...728...63R} {728, 63}

\bibitem[\protect\citeauthoryear{{Roth}, {Kelz}, {Fechner}  et~al.}{{Roth}
  et~al.}{2005}]{roth05}
{Roth} M.~M.,  {Kelz} A.,  {Fechner} T.,   et~al., 2005, \mn@doi [\pasp]
  {10.1086/429877}, \href {http://adsabs.harvard.edu/abs/2005PASP..117..620R}
  {117, 620}

\bibitem[\protect\citeauthoryear{{Roy}, {Kumar}, {Benetti}  et~al.}{{Roy}
  et~al.}{2011}]{roy11}
{Roy} R.,  {Kumar} B.,  {Benetti} S.,   et~al., 2011, \mn@doi [\apj]
  {10.1088/0004-637X/736/2/76}, \href
  {http://adsabs.harvard.edu/abs/2011ApJ...736...76R} {736, 76}

\bibitem[\protect\citeauthoryear{{Rubin}, {Gal-Yam}, {De Cia}  et~al.}{{Rubin}
  et~al.}{2016}]{rubin16b}
{Rubin} A.,  {Gal-Yam} A.,  {De Cia} A.,   et~al., 2016, \mn@doi [\apj]
  {10.3847/0004-637X/820/1/33}, \href
  {http://adsabs.harvard.edu/abs/2016ApJ...820...33R} {820, 33}

\bibitem[\protect\citeauthoryear{{Sahu}, {Anupama}, {Srividya}  \&
  {Muneer}}{{Sahu} et~al.}{2006}]{sahu06}
{Sahu} D.~K.,  {Anupama} G.~C.,  {Srividya} S.,   {Muneer} S.,  2006, \mn@doi
  [\mnras] {10.1111/j.1365-2966.2006.10937.x}, \href
  {http://adsabs.harvard.edu/abs/2006MNRAS.372.1315S} {372, 1315}

\bibitem[\protect\citeauthoryear{{S{\'a}nchez}, {Rosales-Ortega}, {Marino} }{{S{\'a}nchez}
  et~al.}{2012}]{sanchez12}
{S{\'a}nchez} S.~F., {Rosales-Ortega}, F.~F., {Marino}, R.~A.  et~al., 2012, \mn@doi [\aap]
  {10.1051/0004-6361/201219578}, \href
  {http://adsabs.harvard.edu/abs/2012A%26A...546A...2S} {546, A2}

\bibitem[\protect\citeauthoryear{{Sanders}, {Soderberg}, {Gezari}
  et~al.}{{Sanders} et~al.}{2015}]{sanders15}
{Sanders} N.~E.,  {Soderberg} A.~M.,  {Gezari} S.,   et~al., 2015, \mn@doi
  [\apj] {10.1088/0004-637X/799/2/208}, \href
  {http://adsabs.harvard.edu/abs/2015ApJ...799..208S} {799, 208}

\bibitem[\protect\citeauthoryear{{Schlafly} \& {Finkbeiner}}{{Schlafly} \&
  {Finkbeiner}}{2011}]{schlafly11}
{Schlafly} E.~F.,  {Finkbeiner} D.~P.,  2011, \mn@doi [\apj]
  {10.1088/0004-637X/737/2/103}, \href
  {http://adsabs.harvard.edu/abs/2011ApJ...737..103S} {737, 103}

\bibitem[\protect\citeauthoryear{{Schlegel}}{{Schlegel}}{1990}]{sch90}
{Schlegel} E.~M.,  1990, \mnras, \href
  {http://adsabs.harvard.edu/abs/1990MNRAS.244..269S} {244, 269}

\bibitem[\protect\citeauthoryear{{Shivvers}, {Groh}, {Mauerhan}, {Fox}, {Leonard} \&
  {Filippenko}}{{Shivvers} et~al.}{2015}]{shivvers15}
{Shivvers}, I. and {Groh}, J.~H. and {Mauerhan}, J.~C. and {Fox}, O.~D. and 
	{Leonard}, D.~C. and {Filippenko}, A.~V.,  2015, \mn@doi
  [\apj] {10.1088/0004-637X/806/2/213}, \href
  {http://adsabs.harvard.edu/abs/2015ApJ...806..213S} {806, 213}

\bibitem[\protect\citeauthoryear{{Shivvers}, {Zheng}, {Yuk}  \&
  {Filippenko}}{{Shivvers} et~al.}{2016}]{shivvers16}
{Shivvers} I.,  {Zheng} W.,  {Yuk} H.,   {Filippenko} A.~V.,  2016, The
  Astronomer's Telegram, \href
  {http://adsabs.harvard.edu/abs/2016ATel.9381....1S} {9381}

\bibitem[\protect\citeauthoryear{{Silverman}, {Foley}, {Filippenko}
  et~al.}{{Silverman} et~al.}{2012}]{silverman12}
{Silverman} J.~M.,  {Foley} R.~J.,  {Filippenko} A.~V.,   et~al., 2012, \mn@doi
  [\mnras] {10.1111/j.1365-2966.2012.21270.x}, \href
  {http://adsabs.harvard.edu/abs/2012MNRAS.425.1789S} {425, 1789}

\bibitem[\protect\citeauthoryear{{Smartt}}{{Smartt}}{2009}]{smartt09b}
{Smartt} S.~J.,  2009, \mn@doi [\araa] {10.1146/annurev-astro-082708-101737},
  \href {http://adsabs.harvard.edu/abs/2009ARA%26A..47...63S} {47, 63}

\bibitem[\protect\citeauthoryear{{Smartt}, {Eldridge}, {Crockett}  \&
  {Maund}}{{Smartt} et~al.}{2009}]{smartt09a}
{Smartt} S.~J.,  {Eldridge} J.~J.,  {Crockett} R.~M.,   {Maund} J.~R.,  2009,
  \mn@doi [\mnras] {10.1111/j.1365-2966.2009.14506.x}, \href
  {http://adsabs.harvard.edu/abs/2009MNRAS.395.1409S} {395, 1409}

\bibitem[\protect\citeauthoryear{{Stasi{\'n}ska}, {Mateus}, {Sodr{\'e}}  \&
  {Szczerba}}{{Stasi{\'n}ska} et~al.}{2004}]{stasinska04}
{Stasi{\'n}ska} G.,  {Mateus} Jr. A.,  {Sodr{\'e}} Jr. L.,   {Szczerba} R.,
  2004, \mn@doi [\aap] {10.1051/0004-6361:20040117}, \href
  {http://adsabs.harvard.edu/abs/2004A%26A...420..475S} {420, 475}

\bibitem[\protect\citeauthoryear{{Stetson}}{{Stetson}}{1987}]{stetson87}
{Stetson} P.~B.,  1987, \mn@doi [\pasp] {10.1086/131977}, \href
  {http://adsabs.harvard.edu/abs/1987PASP...99..191S} {99, 191}

\bibitem[\protect\citeauthoryear{{Taddia}, {Moquist}, {Sollerman}
  et~al.}{{Taddia} et~al.}{2016}]{taddia16a}
{Taddia} F.,  {Moquist} P.,  {Sollerman} J.,   et~al., 2016, \mn@doi [\aap]
  {10.1051/0004-6361/201527983}, \href
  {http://adsabs.harvard.edu/abs/2016A%26A...587L...7T} {587, L7}

\bibitem[\protect\citeauthoryear{{Turatto}, {Benetti}  \&
  {Cappellaro}}{{Turatto} et~al.}{2003}]{tur03}
{Turatto} M.,  {Benetti} S.,   {Cappellaro} E.,  2003, in {Hillebrandt} W.,
  {Leibundgut} B.,  eds, From Twilight to Highlight: The Physics of Supernovae.
  p.~200 (\mn@eprint {} {arXiv:astro-ph/0211219}), \mn@doi{10.1007/1082854926}

\bibitem[\protect\citeauthoryear{{Valenti}, {Sand}, {Pastorello}
  et~al.}{{Valenti} et~al.}{2014}]{valenti14}
{Valenti} S.,  {Sand} D.,  {Pastorello} A.,   et~al., 2014, \mn@doi [\mnras]
  {10.1093/mnrasl/slt171}, \href
  {http://adsabs.harvard.edu/abs/2014MNRAS.438L.101V} {438, L101}

\bibitem[\protect\citeauthoryear{{Valenti}, {Sand}, {Stritzinger}
  et~al.}{{Valenti} et~al.}{2015}]{valenti15}
{Valenti} S.,  {Sand} D.,  {Stritzinger} M.,   et~al., 2015, \mn@doi [\mnras]
  {10.1093/mnras/stv208}, \href
  {http://adsabs.harvard.edu/abs/2015MNRAS.448.2608V} {448, 2608}

\bibitem[\protect\citeauthoryear{{Valenti}, {Howell}, {Stritzinger}
  et~al.}{{Valenti} et~al.}{2016}]{valenti16}
{Valenti} S.,  {Howell} D.~A.,  {Stritzinger} M.~D.,   et~al., 2016, \mn@doi
  [\mnras] {10.1093/mnras/stw870}, \href
  {http://adsabs.harvard.edu/abs/2016MNRAS.459.3939V} {459, 3939}

\bibitem[\protect\citeauthoryear{{Van Dyk}, {Li}  \& {Filippenko}}{{Van Dyk}
  et~al.}{2003}]{vandyk03}
{Van Dyk} S.~D.,  {Li} W.,   {Filippenko} A.~V.,  2003, \mn@doi [\pasp]
  {10.1086/378308}, \href {http://adsabs.harvard.edu/abs/2003PASP..115.1289V}
  {115, 1289}

\bibitem[\protect\citeauthoryear{{Verheijen}, {Bershady}, {Andersen},
  {Swaters}, {Westfall}, {Kelz}  \& {Roth}}{{Verheijen}
  et~al.}{2004}]{verheijen04}
{Verheijen} M.~A.~W.,  {Bershady} M.~A.,  {Andersen} D.~R.,  {Swaters} R.~A.,
  {Westfall} K.,  {Kelz} A.,   {Roth} M.~M.,  2004, \mn@doi [Astronomische
  Nachrichten] {10.1002/asna.200310197}, \href
  {http://adsabs.harvard.edu/abs/2004AN....325..151V} {325, 151}

\bibitem[\protect\citeauthoryear{{Waxman}, {M{\'e}sz{\'a}ros}  \&
  {Campana}}{{Waxman} et~al.}{2007}]{waxman07}
{Waxman} E.,  {M{\'e}sz{\'a}ros} P.,   {Campana} S.,  2007, \mn@doi [\apj]
  {10.1086/520715}, \href {http://adsabs.harvard.edu/abs/2007ApJ...667..351W}
  {667, 351}

\bibitem[\protect\citeauthoryear{{Weiler}, {Sramek}, {Panagia}, {van der Hulst}
   \& {Salvati}}{{Weiler} et~al.}{1986}]{weiler86}
{Weiler} K.~W.,  {Sramek} R.~A.,  {Panagia} N.,  {van der Hulst} J.~M.,
  {Salvati} M.,  1986, \mn@doi [\apj] {10.1086/163944}, \href
  {http://adsabs.harvard.edu/abs/1986ApJ...301..790W} {301, 790}

\bibitem[\protect\citeauthoryear{{Woosley}, {Pinto}, {Martin}  \&
  {Weaver}}{{Woosley} et~al.}{1987}]{woosley87}
{Woosley} S.~E.,  {Pinto} P.~A.,  {Martin} P.~G.,   {Weaver} T.~A.,  1987,
  \mn@doi [\apj] {10.1086/165402}, \href
  {http://adsabs.harvard.edu/abs/1987ApJ...318..664W} {318, 664}

\bibitem[\protect\citeauthoryear{{Yaron}, {Perley}, {Gal-Yam}  et~al.}{{Yaron}
  et~al.}{2017}]{yaron17}
{Yaron} O.,  {Perley} D.~A.,  {Gal-Yam} A.,   et~al., 2017, \mn@doi [Nature
  Physics] {10.1038/nphys4025}, \href
  {http://adsabs.harvard.edu/abs/2017NatPh..13..510Y} {13, 510}

\bibitem[\protect\citeauthoryear{{Yuan}, {Jerkstrand}, {Valenti}
  et~al.}{{Yuan} et~al.}{2016}]{yuan16}
{Yuan} F.,  {Jerkstrand} A.,  {Valenti} S.,   et~al., 2016, \mn@doi [\mnras]
  {10.1093/mnras/stw1419}, \href
  {http://adsabs.harvard.edu/abs/2016MNRAS.461.2003Y} {461, 2003}

\makeatother
\end{thebibliography}
\end{document}